\documentclass[11pt]{article}
\pdfoutput=1
\usepackage{amsmath,amssymb,amsthm,amsxtra,overpic,bbm,bm,epsfig,slashed}
\usepackage{color,subfigure}
\usepackage[square,sort,comma,numbers]{natbib}
\def\thefootnote{\fnsymbol{footnote}}
\allowdisplaybreaks[4]

\def\thefootnote{\fnsymbol{footnote}}

\newcommand{\bq}{\begin{eqnarray}}
\newcommand{\nq}{\end{eqnarray}}

\newcommand{\LH}{\text{L}}
\newcommand{\RH}{\text{R}}

\newcommand{\ini}{{\bf i}}
\newcommand{\fin}{{\bf f}}

\def\dx{\mathrm{d}x}
\def\dy{\mathrm{d}y}
\def\dz{\mathrm{d}z}

\begin{document}

\hfill {\tt IPPP/19/81}

\begin{center}
{\Large\bf $CP$ violation and circular polarisation in neutrino radiative decay}
\end{center} 
\vspace{0.2cm}

\begin{center}
{\bf Shyam Balaji,$^1$}\footnote{Email: \tt shyam.balaji@sydney.edu.au} ~
{\bf Maura Ramirez-Quezada$^2$}\footnote{Email: \tt maura.e.ramirez-quezada@durham.ac.uk} ~
and 
{\bf Ye-Ling Zhou$^3$}\footnote{Email: \tt ye-ling.zhou@soton.ac.uk}
\\\vspace{5mm}
{$^1$ School of Physics, The  University of Sydney, NSW 2006, Australia} \\
{$^2$ Institute for Particle Physics Phenomenology, Department of Physics, \\ Durham University, Durham DH1 3LE, United Kingdom} \\
{$^3$ School of Physics and Astronomy, University of Southampton,\\
SO17 1BJ Southampton, United Kingdom } 
\\
\end{center}

\vspace{1.5cm} 

\begin{abstract}

The radiative decay of neutral fermions has been studied for decades but $CP$ violation induced within such a paradigm has evaded attention. $CP$ violation in these processes can produce an asymmetry between circularly polarised directions of the radiated photons and produces an important source of net circular polarisation in particle and astroparticle physics observables. The results presented in this work outlines the general connection between $CP$ violation and circular polarisation for both Dirac and Majorana fermions and can be used for any class of models that produce such radiative decays. The total $CP$ violation is calculated based on a widely studied Yukawa interaction considered in both active and sterile neutrino radiative decay scenarios as well as searches for dark matter via direct detection and collider signatures. Finally, the phenomenological implications of the formalism on keV sterile neutrino decay, leptogenesis-induced right-handed neutrino radiative decay and IceCube-driven heavy dark matter decay are discussed.

\end{abstract}
\begin{flushleft}
\hspace{0.8cm} PACS number(s): \\
\hspace{0.8cm} Keywords: $CP$ violation, circular polarisation, neutrino radiative decay, dark matter, leptogenesis
\end{flushleft}

\def\thefootnote{\arabic{footnote}}
\setcounter{footnote}{0}

\newpage

\section{Introduction}

For decades, studies of neutrinos have deepened our understanding of nature \cite{Tanabashi:2018oca}. Although their very small but non-zero masses (for at least two of their generations) and lepton flavour mixing have been observed and verified by neutrino oscillation experiments, some fundamental questions about neutrinos such as their electromagnetic properties, $CP$ violation, whether they are Dirac or Majorana fermions and if they have additional species existing in nature remain unknown.  

The studies of neutrino radiative decays dates back fourty years \cite{Shrock:1974nd, Petcov:1976ff, Goldman:1977jx} and beyond. Assuming neutrinos are electrically neutral fermions (Dirac or Majorana), their electromagnetic dipole moments (EDMs) can be generated at various loop levels and neutrino radiative decays $\nu_\ini \to \nu_\fin + \gamma$ are induced by off-diagonal parts of the EDMs \cite{Schechter:1981hw,Pal:1981rm,Schechter:1981cv,Shrock:1982sc,Nieves:1981zt,Kayser:1982br}. Charged current interaction contributions in the Standard Model (SM) have previously been calculated at one-loop level in \cite{Schechter:1981hw,Pal:1981rm,Schechter:1981cv,Shrock:1982sc,Nieves:1981zt} and later studied in detail in \cite{Dvornikov:2003js, Dvornikov:2004sj}. However, these contributions are tiny due to the large mass hierarchy between the active neutrinos and the $W$ boson as there is currently no positive experimental indication in favour of their existence. Neutrino electromagnetic interactions therefore provide a tantalising probe for new physics (NP) beyond the SM (see \cite{Giunti:2014ixa} for a comprehensive review). 

If more massive neutrinos exist, then these heavy neutrinos may decay to the lighter active neutrinos radiatively. These heavier neutrinos will consequently have a larger decay width due to the existence of such decay channels. Various hypothetical heavier neutrinos have been historically introduced, motivated by a combination of theoretical and phenomenological reasons. Some of the most famous ones are those introduced in the type-I seesaw mechanism \cite{Minkowski:1977sc, Yanagida:1979as, GellMann:1980vs, Glashow:1979nm, Mohapatra:1979ia, Schechter:1980gr}, 
which was proposed in order to address the origin of sub-eV left-handed neutrino masses. 
Phenomenological motivations have suggested keV sterile neutrinos as dark matter (DM) candidates  
to explain the detection of a 3.5 keV X-ray line in \cite{Bulbul:2014sua,Boyarsky:2014jta} (for some representative reviews, see \cite{Gariazzo:2015rra,Adhikari:2016bei,Xing:2019vks}). 
Very heavy DM was also proposed 
\cite{Chianese:2016opp,Aartsen:2018mxl} in order to explain the IceCube data \cite{Aartsen:2013bka,Aartsen:2016xlq}. 
Radiative decays of such heavy particles may be more significant than those of active neutrinos due to their very large relative mass. Hence, radiative decay is typically a major channel of importance in detecting possible keV sterile neutrino DM. 

$CP$ violation may exist in various processes involving neutrinos. At low energy, neutrino oscillations provide the best way to clarify its existence in the neutrino sector. Combined analysis of current accelerator neutrino oscillation data \cite{Abe:2018wpn} supports large $CP$ violation in the appearance channel of neutrino oscillations \cite{Abe:2017uxa, Adamson:2017gxd}. The next-generation neutrino oscillation experiments DUNE and T2HK are projected to observe $CP$ violation in the near future \cite{Abi:2018alz,Abe:2015zbg,Abe:2016ero}.  
At high energy, the most well-studied process involving $CP$ violation is the very heavy right-handed neutrino decaying into  SM leptons and the Higgs boson. This effect is the source of the so-called thermal leptogenesis phenomenon, which can explain the observed matter-antimatter asymmetry in our Universe  \cite{Fukugita:1986hr}. On the other hand, if these heavy neutrinos have lighter masses, specifically around the GeV scale, $CP$ violation may appear in right-handed neutrino oscillations, which provides an alternative mechanism for leptogenesis \cite{Akhmedov:1998qx} (See \cite{Buchmuller:2004nz,Davidson:2008bu} for some reviews). 

In this work we study $CP$ violation in radiative decays of both Dirac and Majorana neutrinos. 
Whilst neutrino radiative decays have been extensively studied for some mass regions of neutrinos, $CP$ violation in these processes has not been studied for a more general spectrum of mass scales with very few exceptions e.g. \cite{Bell:2008fm}.
Recently, it was suggested in \cite{Boehm:2017nrl} that a net circular polarisation, specifically an asymmetry between two circularly polarised photons $\gamma_+$ and $\gamma_-$, can be generated if $CP$ is violated in neutrino radiative decays. Therefore, the circular polarisation of photons provides a potentially crucial probe to prove the existence of $CP$ violation in the neutrino and DM sectors.

This work builds a formulation to describe both $CP$ violation in neutrino radiative decays and also the resulting asymmetry between the produced photons $\gamma_+$ and $\gamma_-$. In Section~\ref{sec:framework}, we outline the most general formalism of $CP$ violation and circular polarisation in terms of form factors where the result is independent of the neutrino model or mass scale. In Section~\ref{sec:CP}, we discuss $CP$ violation based on a simplified neutrino model. We begin this section with a discussion about the size of $CP$ asymmetry for the SM contribution and then consider how $CP$ violation can be enhanced via new interactions. A comprehensive analytical calculation of $CP$ asymmetry based on Yukawa type NP interactions is then performed in Section \ref{sec:new_int}, this type of simple interaction has a wide ensemble of phenomenological applications which is shown in Section~\ref{sec:applications}. Finally, we summarise our results in Section~\ref{sec:conclusion}.

\section{The framework\label{sec:framework}}

In this section we shall set up the framework for computation of $CP$ violation in neutrino radiative decays and the general connection with circular polarisation generated by such processes. Discussion in this section is fully independent of neutrino interactions and thus is applicable to  any other electrically neutral fermion with mass at any scale. 

Discussions in Section \ref{subsec:matrix} and \ref{subsec:CP} assume neutrinos are Dirac fermions. The extension to Majorana neutrinos will be given in Section \ref{subsec:decay}. 

\subsection{Matrix element for polarised particles\label{subsec:matrix}}

Assuming fermions are Dirac particles, the amplitude for the process $\nu_\ini \to \nu_\fin + \gamma_{\pm}$ is given by
\begin{eqnarray}
i \mathcal{M} (\nu_\ini \to \nu_\fin + \gamma_{\pm})  = i \bar{u}(p_\fin) \Gamma_{\fin \ini }^\mu(q^2) u(p_\ini) \varepsilon^*_{\pm,\mu}(q) \,. \label{eq:decay_amplitude}
\end{eqnarray}
Here, $u(p_\ini)$ and $u(p_\fin)$ are spinors for the initial $\nu_\ini$ and final $\nu_\fin$ state neutrinos respectively. By momentum conservation, the photon  momentum is $q=p_\ini - p_\fin$. The spinors include the spin polarisation of the fermions, this will be discussed in more detail in the next subsection in a specified inertial reference frame. The transition form factor is then parametrised as per \cite{Nieves:1981zt, Shrock:1982sc, Kayser:1982br, Kayser:1984ge}
\begin{eqnarray}
\Gamma^\mu_{\fin \ini } (q^2) &=& f^\text{Q}_{\fin \ini } (q^2) \gamma^\mu -
f^\text{M}_{\fin \ini } (q^2) i \sigma^{\mu \nu} q_\nu +
f^\text{E}_{\fin \ini } (q^2) \sigma^{\mu \nu} q_\nu \gamma_5 +
f^\text{A}_{\fin \ini } (q^2) ( q^2 \gamma^\mu - q^\mu \slashed{q})
\gamma_5 \,.
\end{eqnarray}
We will not consider electrically charged neutrinos, namely we require that $f^Q=0$. The modification to the result in the case of non-zero $f^Q$ will be mentioned at the end of this section. By requiring the photon to be on-shell $q^2=0$ and choosing the Lorenz gauge $q\cdot \varepsilon_p = 0$, the anapole does not contribute. In this case, only the electromagnetic dipole moment contributes to the neutrino radiative decay. We then rewrite the form factor as 
\begin{eqnarray}
\Gamma^\mu_{\fin \ini } (q^2) &=& i \sigma^{\mu \nu} q_\nu [f^\text{L}_{\fin \ini } (q^2) P_{\text{L}} + f^\text{R}_{\fin \ini } (q^2) P_{\text{R}}] \,, 
\end{eqnarray}
where $f^{\text{L,R}}_{\fin \ini} = - f^\text{M}_{\fin \ini} \pm i f^\text{E}_{\fin \ini}$ and the chiral projection operators are defined as $P_{\text{L,R}} = \frac{1}{2}(1 \mp \gamma_5)$. 
The decay widths for $\nu_\ini \to \nu_\fin+ \gamma_{\pm}$ are then given by
\begin{eqnarray} 
\label{eq:Gamma}
\Gamma(\nu_\ini \to \nu_\fin+ \gamma_\pm)
&=&\frac{m_\ini^2-m_\fin^2}{16\pi m^3_\ini} |\mathcal{M} (\nu_\ini \to \nu_\fin + \gamma_{\pm})|^2\,.
\end{eqnarray}
The amplitudes $\mathcal{M}(\nu_{\ini} \to \nu_{\fin} + \gamma_\pm)$ are directly correlated with the coefficients 
\begin{eqnarray} \label{eq:amplitudes_1}
\mathcal{M}(\nu_{\ini} \to \nu_{\fin} + \gamma_+) &=& +\sqrt{2} f^{\text{L}}_{\fin \ini } (m_\ini^2 - m_\fin^2) \,, \nonumber\\
\mathcal{M}(\nu_{\ini} \to \nu_{\fin} + \gamma_-) &=& - \sqrt{2} f^{\text{R}}_{\fin \ini } (m_\ini^2 - m_\fin^2) \,. 
\end{eqnarray}
which are derived in detail in Appendix~\ref{sec:rest_frame}. 
The sum of the decay widths for $\nu_\ini \to \nu_\fin+ \gamma_+$ and $\nu_\ini \to \nu_\fin+ \gamma_-$ yields the total radiative decay width $\Gamma(\nu_\ini \to \nu_\fin + \gamma)$. 

Again, if we only consider radiative decay for an electrically neutral antineutrino, the amplitudes of radiative decay $\bar{\nu}_\ini \to \bar{\nu}_\fin + \gamma_\pm$ are then given by 
\begin{eqnarray}
i \mathcal{M} (\bar{\nu}_\ini \to \bar{\nu}_\fin + \gamma_{\pm}) 
= i \bar{v}(p_\ini) \bar{\Gamma}_{\ini \fin}^\mu(q^2) v(p_\fin) \varepsilon^*_{\pm,\mu}(q) \,,
\end{eqnarray}
where $v(p_\ini)$ and $v(p_\fin)$ are antineutrino spinors. The decay width for $\bar{\nu}_\ini \to \bar{\nu}_{\fin,s'}+ \gamma_l$ is
\begin{eqnarray}
\Gamma(\bar{\nu}_\ini \to \bar{\nu}_\fin+ \gamma_\pm)
&=&\frac{m_\ini^2-m_\fin^2}{16\pi m_\ini^3} |\mathcal{M} (\bar{\nu}_\ini \to \bar{\nu}_\fin + \gamma_{\pm})|^2 \,.
\end{eqnarray}
By parametrising the form factor in a similar form to before, we have
\begin{eqnarray}
\bar{\Gamma}^\mu_{\ini \fin} (q^2) &=& i \sigma^{\mu \nu} q_\nu [\bar{f}^\text{L}_{\ini \fin} (q^2) P_{\text{L}} + \bar{f}^\text{R}_{\ini \fin} (q^2) P_{\text{R}}] \, ,
\end{eqnarray}
with $\bar{f}^{\text{L,R}}_{\ini \fin} =- \bar{f}^\text{M}_{\ini \fin} \pm i \bar{f}^\text{E}_{\ini \fin}$. Therefore,  the amplitudes can be written in a similar fashion following Eq.~\eqref{eq:amplitudes_1}, i.e. by replacing $f^{\text{L}}_{\fin \ini }$ and $f^{\text{R}}_{\fin \ini }$ by $\bar{f}^{\text{L}}_{\ini \fin}$ and $\bar{f}^{\text{R}}_{\ini \fin}$ respectively (see the proof in Appendix~\eqref{sec:rest_frame}). 
These formulae can be further simplified with the help of the $CPT$ theorem, which is satisfied in all Lorentz invariant local quantum field theories with a Hermitian Hamiltonian. 
Due to $CPT$ invariance, $\bar{\nu}_\ini \to \bar{\nu}_\fin + \gamma_{\mp}$ and $\nu_\fin + \gamma_{\pm} \to \nu_\ini $ have the same amplitude, and thus $\bar{f}^\text{M,E}_{\ini \fin} (q^2) = - f^\text{M,E}_{\ini \fin} (q^2)$ is satisfied \cite{Giunti:2014ixa}, leading to 
\begin{eqnarray} \label{eq:CPT}
 \bar{f}^\text{L}_{\ini \fin} (q^2) = - f^\text{L}_{ \ini \fin} (q^2) \,, \quad
 \bar{f}^\text{R}_{\ini \fin} (q^2) = - f^\text{R}_{ \ini \fin} (q^2) \,.
\end{eqnarray} 
Hence, amplitudes $\mathcal{M}(\bar{\nu}_{\ini} \to \bar{\nu}_{\fin} + \gamma_+)$  can be simplified to 
\begin{eqnarray} \label{eq:amplitudes_2}
\mathcal{M}(\bar{\nu}_{\ini} \to \bar{\nu}_{\fin} + \gamma_+) &=& +\sqrt{2} f^{\text{L}}_{\ini \fin} (m_\ini^2 - m_\fin^2) \,, \nonumber\\
\mathcal{M}(\bar{\nu}_{\ini} \to \bar{\nu}_{\fin} + \gamma_-) &=& -\sqrt{2} f^{\text{R}}_{\ini \fin} (m_\ini^2 - m_\fin^2) \,.  
\end{eqnarray}

Physical neutrinos and antineutrinos are related by a $CP$ transformation which interchanges particles with antiparticles and replaces momentum by its parity conjugate $\tilde{p}=(p_0,-\vec{p})$. The $CP$ transformation reverses the momentum but preserves angular momentum. As a consequence, the polarisation is reversed. Performing a $CP$ transformation for $\nu_\ini(p_\ini) \to \nu_\fin(p_\fin) + \gamma_{\pm}(q)$ gives rise to antineutrino channels with reversed 3D momentum and reversed photon polarisations in the final states $\bar{\nu}_\ini(\tilde{p}_\ini) \to \bar{\nu}_\fin(\tilde{p}_\fin) + \gamma_{\mp}(\tilde{q})$. Since the amplitude is parity-invariant, the amplitude of the  process is equivalent to $\bar{\nu}_\ini(p_\ini) \to \bar{\nu}_\fin(p_\fin) + \gamma_{\mp}(q)$. Therefore, the radiative decay of  antineutrinos can be represented as a $CP$ conjugate of the decay of neutrinos
\begin{eqnarray}
i \mathcal{M} (\bar{\nu}_\ini \to \bar{\nu}_\fin + \gamma_{\pm}) 
= i \mathcal{M}^{CP} (\nu_\ini \to \nu_\fin + \gamma_{\mp}) \,.
\end{eqnarray}
In the case of $CP$ conservation, both $f^\text{E}_{ \ini \fin} (q^2)$ and $f^\text{M}_{ \ini \fin} (q^2)$ are Hermitian i.e. $f^\text{M,E}_{ \ini \fin} (q^2) = [f^\text{M,E}_{ \fin \ini} (q^2)]^*$. This leads to  $f^\text{L,R}_{ \ini \fin} (q^2) = [f^\text{R,L}_{ \fin \ini} (q^2)]^*$, namely,
$\bar{f}^\text{L,R}_{\ini \fin} (q^2) = - [f^\text{R,L}_{\fin \ini } (q^2)]^*$ \cite{Broggini:2012df,Giunti:2014ixa}. And eventually, we arrive at the identity 
\begin{eqnarray}
\Gamma(\nu_\ini \to \nu_\fin+ \gamma_\pm) - \Gamma(\bar{\nu}_\ini \to \bar{\nu}_\fin+ \gamma_\mp) \propto
|\mathcal{M} (\nu_\ini \to \nu_\fin + \gamma_{\pm})|^2 - |\mathcal{M}^{CP} (\nu_\ini \to \nu_\fin + \gamma_{\pm})|^2 = 0 \,.
\end{eqnarray}
However, a $CP$ violating source in the interaction may contribute at loop level and break this equality.

\subsection{Correlation between $CP$ asymmetry and circular polarisation\label{subsec:CP}} 

We define the $CP$ asymmetry between the radiative decay $\nu_\ini \to \nu_\fin+ \gamma_+$ and its $CP$ conjugate process $\bar{\nu}_\ini \to \bar{\nu}_\fin+ \gamma_-$ as 
\begin{eqnarray}
\Delta_{CP,+} &=& \frac{\Gamma(\nu_\ini \to \nu_\fin+ \gamma_+) - 
\Gamma(\bar{\nu}_\ini \to \bar{\nu}_\fin+ \gamma_-)}
{\Gamma(\nu_\ini \to \nu_\fin+ \gamma) + 
\Gamma(\bar{\nu}_\ini \to \bar{\nu}_\fin+ \gamma)} \,.
\end{eqnarray}
The $CP$ asymmetry between $\nu_\ini \to \nu_\fin+ \gamma_-$ and its $CP$ conjugate process $\bar{\nu}_\ini \to \bar{\nu}_\fin+ \gamma_+$, $\Delta_{CP,-}$, is defined by exchanging $+$ and $-$ signs.
The photon polarisation independent $CP$ asymmetry is obtained by summing $\Delta_{CP,+}$ and $\Delta_{CP,-}$ together which yields
\begin{eqnarray}
\Delta_{CP} &=& \frac{\Gamma(\nu_\ini \to \nu_\fin+ \gamma_{+}) - \Gamma(\bar{\nu}_\ini \to \bar{\nu}_\fin+ \gamma_{-}) + \Gamma(\nu_\ini \to \nu_\fin+ \gamma_{-})
- \Gamma(\bar{\nu}_\ini \to \bar{\nu}_\fin+ \gamma_{+}) }
{\Gamma(\nu_\ini \to \nu_\fin+ \gamma) + 
\Gamma(\bar{\nu}_\ini \to \bar{\nu}_\fin+ \gamma)} \,.
\end{eqnarray}
It is also convenient to define the asymmetry between the radiated photons $\gamma_+$ and $\gamma_-$ as
\begin{eqnarray}
\Delta_{+-}&=& \frac{\Gamma(\nu_\ini \to \nu_\fin+ \gamma_{+}) + \Gamma(\bar{\nu}_\ini \to \bar{\nu}_\fin+ \gamma_{+}) 
- \Gamma(\nu_\ini \to \nu_\fin+ \gamma_{-}) - \Gamma(\bar{\nu}_\ini \to \bar{\nu}_\fin+ \gamma_{-})}
{\Gamma(\nu_\ini \to \nu_\fin+ \gamma) + 
\Gamma(\bar{\nu}_\ini \to \bar{\nu}_\fin+ \gamma)} \,.
\end{eqnarray}
Given equal numbers for initial neutrinos and antineutrinos, $\Delta_{+-}$ represents the fraction $({\rm N}_{\gamma_+} - {\rm N}_{\gamma_-}) / ({\rm N}_{\gamma_+} + {\rm N}_{\gamma_-})$, where ${\rm N}_{\gamma_+}$ and ${\rm N}_{\gamma_-} $ are the number of polarised photons $\gamma_+$ and $\gamma_-$ produced by the radiative decays respectively.  It is this source that generates circular polarisation for the radiated photons giving rise to a non-zero Stokes parameter $V$. 

Therefore, a non-zero $\Delta_{+-}$ is a source of circular polarisation for the photon produced by the radiative decay. 
Since the phase spaces are the same for neutrino and antineutrino channels, these formulae can be simplified to
\begin{eqnarray} \label{eq:CP_v1}
\Delta_{CP,+} &=& \frac{|f^{\text{L}}_{\fin \ini }|^2 - |f^{\text{R}}_{\ini \fin}|^2}
{|f^{\text{L}}_{\fin \ini }|^2 + |f^{\text{R}}_{\fin \ini }|^2 + |f^{\text{R}}_{\ini \fin}|^2 + |f^{\text{L}}_{\ini \fin}|^2} \,,\nonumber\\
\Delta_{CP,-} &=& \frac{|f^{\text{R}}_{\fin \ini }|^2 - |f^{\text{L}}_{\ini \fin}|^2}
{|f^{\text{L}}_{\fin \ini }|^2 + |f^{\text{R}}_{\fin \ini }|^2 + |f^{\text{R}}_{\ini \fin}|^2 + |f^{\text{L}}_{\ini \fin}|^2} \,,
\end{eqnarray}
as well as
\begin{eqnarray} \label{eq:CP_v2}
\Delta_{CP} &=& \frac{|f^{\text{L}}_{\fin \ini }|^2 + |f^{\text{R}}_{\fin \ini }|^2 - |f^{\text{R}}_{\ini \fin}|^2 - |f^{\text{L}}_{\ini \fin}|^2}
{|f^{\text{L}}_{\fin \ini }|^2 + |f^{\text{R}}_{\fin \ini }|^2 + |f^{\text{R}}_{\ini \fin}|^2 + |f^{\text{L}}_{\ini \fin}|^2} \,,\nonumber\\ 
\Delta_{+-}&=& \frac{|f^{\text{L}}_{\fin \ini }|^2 - |f^{\text{R}}_{\fin \ini }|^2 - |f^{\text{R}}_{\ini \fin}|^2 + |f^{\text{L}}_{\ini \fin}|^2}
{|f^{\text{L}}_{\fin \ini }|^2 + |f^{\text{R}}_{\fin \ini }|^2 + |f^{\text{R}}_{\ini \fin}|^2 + |f^{\text{L}}_{\ini \fin}|^2} \,.
\end{eqnarray}
The total $CP$ asymmetry and the asymmetry between $\gamma_+$ and $\gamma_-$ follows simple relations with $\Delta_{CP,+}$ and $\Delta_{CP,-}$ as  
\begin{eqnarray} \label{eq:CP&CP}
\Delta_{CP} &=& \Delta_{CP,+} + \Delta_{CP,-} \,,\nonumber\\ 
\Delta_{+-}&=& \Delta_{CP,+} - \Delta_{CP,-} \,.
\end{eqnarray}

Therefore, we arrive at an important result that the generation of circular polarisation is essentially dependent upon 
$CP$ asymmetry between neutrino radiative decay and its $CP$ conjugate process. Note that we have not included any details related to the Lagrangian or interactions yet. Given any neutral fermion, its radiative decay can always be parametrised by the electromagnetic dipole moments with coefficients $f_{\fin \ini}^{\LH}$ and $f_{\fin \ini}^{\RH}$ (as well as $\bar{f}_{\ini \fin}^{\LH}$ and $\bar{f}_{\ini \fin}^{\RH}$ for its antiparticle), we then arrive at the correlations between $CP$ violation and circular polarisation in Eq.~\eqref{eq:CP&CP} with their definitions in Eqs.~\eqref{eq:CP_v1} and \eqref{eq:CP_v2}.

Another source of asymmetry between polarised photons is the existence of an initial number asymmetry between neutrinos and antineutrinos \cite{Boehm:2017nrl}. There may be some other $CP$ violating sources in particle physics which can induce this condition \cite{Gorbunov:2016zxf}. On the other hand, this kind of asymmetry is more likely to be generated in extreme astrophysical environments. For example, in supernovae explosions, the asymmetry between sterile neutrinos and antineutrinos may be generated because of the different matter effects during neutrino and antineutrino propagation \cite{Shi:1993ee,Raffelt:2011nc}. In the rest of this paper, we will only consider circular polarisation directly produced by the $CP$ violating decays between neutrinos and antineutrinos.

Now we may turn our attention to obtaining non-zero $CP$ violation for the radiative decay. 
For $\nu_\ini \to \nu_\fin+ \gamma_{+}$ and $\nu_\ini \to \nu_\fin+ \gamma_{-}$, we parametrise
the effective coefficients $f^{\text{L}}_{\fin \ini }$ and $f^{\text{R}}_{\fin \ini }$, these should be obtained from the relevant loop calculations in the form 
\begin{eqnarray} \label{eq:CK}
f^{\text{L}}_{\fin \ini } = \sum_l C_{l} K_{l}^{\text{L}}\,, \quad
f^{\text{R}}_{\fin \ini } = \sum_l C_{l} K_{l}^{\text{R}},
\end{eqnarray}
without loss of generality. Here, we have used $l$ to classify the different categories of loop contributions. For each loop category $l$, $C_l$ factorises out all coefficients of operators contributing to the diagram. $K_l^{\text{L}}$ and $K_l^{\text{R}}$ represents the pure loop kinematics after coefficients are extracted out. 
As a consequence, $\bar{f}^{\text{L}}_{\ini \fin}$ and $\bar{f}^{\text{R}}_{\ini \fin}$ (namely $-f^{\text{L}}_{\ini \fin}$ and $-f^{\text{R}}_{\ini \fin}$) corresponding to the effective parameters for $\bar{\nu}_\ini \to \bar{\nu}_\fin + \gamma_\pm$, can always be represented in the form
\footnote{To clarify how this parametrisation is valid, we write out the subscripts explicitly, $f^{\text{L}}_{\fin \ini } = \sum_l (C_{l})_{\fin \ini} (K_{l}^{\text{L}})_{\fin \ini}$ and $f^{\text{R}}_{\fin \ini } = \sum_l (C_{l})_{\fin \ini} (K_{l}^{\text{R}})_{\fin \ini}$. Similarly, we can write out $f^{\text{L}}_{\ini \fin} = \sum_l (C_{l})_{\ini \fin} (K_{l}^{\text{L}})_{\ini \fin}$ and $f^{\text{R}}_{\ini \fin} = \sum_l (C_{l})_{\ini \fin} (K_{l}^{\text{R}})_{\ini \fin}$. One can simplify $f^{\text{L}}_{\ini \fin}$ and $f^{\text{R}}_{\ini \fin}$ in the following steps. 1) The coefficient $(C_{l})_{\ini \fin}$ must be the complex conjugate of $(C_{l})_{\fin \ini}$ since both processes are $CP$ conjugates of one another. 2) $(K_{l}^{\text{L}})_{\ini \fin}$ and $(K_{l}^{\text{R}})_{\ini \fin}$, as pure kinetic terms, must satisfy $T$ parity, namely they must be invariant under the interchange of the initial and final state neutrinos $\nu_\ini \leftrightarrow \nu_\fin$, the chiralities must also be interchanged $\text{L} \leftrightarrow \text{R}$, namely, $(K_{l}^{\text{L}})_{\ini \fin} = (K_{l}^{\text{R}})_{\fin \ini}$ and $(K_{l}^{\text{R}})_{\ini \fin} = (K_{l}^{\text{L}})_{\fin \ini}$. Therefore, $f^{\text{L}}_{\ini \fin}$ and $f^{\text{R}}_{\ini \fin}$ can be re-written to be
$f^{\text{L}}_{\ini \fin} = \sum_l (C_{l})_{\fin \ini}^* (K_{l}^{\text{R}})_{\fin \ini}$ and $f^{\text{R}}_{\ini \fin} = \sum_l (C_{l})_{\fin \ini}^* (K_{l}^{\text{L}})_{\fin \ini}$. 
}

\begin{eqnarray}
f^{\text{L}}_{\ini \fin} = \sum_l C_{l}^* K_{l}^{\text{R}} \,, \quad
f^{\text{R}}_{\ini \fin} = \sum_l C_{l}^* K_{l}^{\text{L}} \,.  
\end{eqnarray}
The $CP$ asymmetries with respect to the photon polarisations can then be simplified to
\begin{eqnarray}
\Delta_{CP,+} \propto 
|f^{\text{L}}_{\fin \ini}| - |f^{\text{R}}_{\ini \fin}| 
&=& - 4 \sum_{l \neq l'}\text{Im}(C_{l} C_{l'}^*) 
\text{Im}( K^{\text{L}}_{l} K^{\text{L}\,*}_{l'}) \,, \nonumber\\
\Delta_{CP,-} \propto 
|f^{\text{R}}_{\fin \ini}| - |\bar{f}^{\text{L}}_{\ini \fin}| 
&=& - 4 \sum_{l \neq l'}\text{Im}(C_{l} C_{l'}^*)
\text{Im} (K^{\text{R}}_{l} K^{\text{R}\,*}_{l'})\,.
\end{eqnarray}
Therefore, a non-zero $CP$ asymmetry is determined by non-vanishing $\text{Im}(C_{l} C_{l'}^*)$ and non-vanishing $\text{Im}(K^{\text{L}}_{l} K^{\text{L}\,*}_{l'})$ (or $\text{Im}(K^{\text{R}}_{l} K^{\text{R}\,*}_{l'})$) from loops $l$ and $l'$. 

While the imaginary part of $\text{Im}(C_{l} C_{l'}^*)$ is straightforwardly obtained from the relevant terms in the Lagrangian, the main task is to compute the imaginary parts of $K^{\text{L}}_l K^{\text{L}\, *}_{l'}$ and $K^{\text{R}}_l K^{\text{L}\, *}_{l'}$. 
In order to achieve non-zero values of these imaginary parts, one may apply the optical theorem which can be expressed as
\begin{eqnarray}
\text{Im}\mathcal{M}(a \to b) = \frac{1}{2}\sum_c \int d \Pi_c \, \mathcal{M}^*(b\to c) \mathcal{M}(a\to c) \,,
\end{eqnarray}
where the sum runs over all possible sets $c$ of final-state particles  \cite{Peskin:1995ev}. 
Fixing $a=\nu_\ini$ and $b=\nu_\fin+\gamma$, $c$ has to include an odd number of fermions plus arbitrary bosons. All particles heavier than $\nu_\ini$ cannot be included in $c$ since this would violate energy-momentum conservation. In the next section, we will explicitly show how to derive a non-zero analytical result for $\text{Im}(K^{\text{R}}_{l} K^{\text{R}\,*}_{l'})$ based on a simplified NP model where $\text{Im}(K^{\text{L}}_{l} K^{\text{L}\,*}_{l'})$ is negligibly small.

\subsection{$CP$ violation in Majorana neutrino radiative decay\label{subsec:decay}}

The above discussion is only limited to Dirac neutrinos. However, neutrinos may also be Majorana particles i.e. where the neutrino is identical to the antineutrino but with potentially different kinematics. In this case, both the neutrino and antineutrino modes must be considered together. 
The amplitude is then given by 
$i \mathcal{M}^\text{M}(\nu_\ini \to \nu_\fin + \gamma_\pm) = i \mathcal{M} (\nu_\ini \to \nu_\fin + \gamma_\pm) + i \mathcal{M} (\bar{\nu}_\ini \to \bar{\nu}_\fin + \gamma_\pm) 
$. 
Taking the explicit formulas for the amplitudes given in Eq~\eqref{eq:amplitudes_1} and \eqref{eq:amplitudes_2}, we obtain results with definite spins in the initial and final states as 
\begin{eqnarray} \label{eq:amplitude_Majorana}
\mathcal{M}^\text{M}(\nu_{\ini} \to \nu_{\fin} + \gamma_+) &=& +
\sqrt{2} [ f^{\text{L}}_{\fin \ini } - f^{\text{L}}_{\ini \fin} ] (m_\ini^2 - m_\fin^2) \,, \nonumber\\
\mathcal{M}^\text{M}(\nu_{\ini} \to \nu_{\fin} + \gamma_-) &=& -
\sqrt{2} [ f^{\text{R}}_{\fin \ini } - f^{\text{R}}_{\ini \fin} ] (m_\ini^2 - m_\fin^2) \,, 
\end{eqnarray}
The decay width $\Gamma^\text{M}(\nu_\ini \to \nu_\fin + \gamma_\pm)$ is still written in the form shown in Eq. ~\eqref{eq:Gamma}. 

For Majorana fermions, the $CP$ violation is identical to that obtained from $P$ violation alone i.e. 
the $CP$ asymmetry is essentially the same as the asymmetry between the two polarised photons $\Delta_{+-}^\text{M}$ 
\begin{eqnarray}
\Delta_{CP,+}^\text{M} = - \Delta_{CP,-}^\text{M} = \Delta_{+-}^\text{M} &=& \frac{\Gamma^\text{M}(\nu_\ini \to \nu_\fin+ \gamma_{+})  
- \Gamma^\text{M}(\nu_\ini \to \nu_\fin+ \gamma_{-}) }
{\Gamma^\text{M}(\nu_\ini \to \nu_\fin+ \gamma)} \,.
\end{eqnarray}
The $CP$ asymmetry without considering the polarisation of the radiated photon is zero, namely, $\Delta_{CP}^\text{M} = \Delta_{CP,+}^\text{M} + \Delta_{CP,-}^\text{M} = 0$. With the help of Eq.~\eqref{eq:amplitude_Majorana}, we can express $\Delta_{+-}^\text{M}$ in the form of electromagnetc dipole parameters as 
\begin{eqnarray} \label{eq:CP_M_v2}
\Delta_{+-}^\text{M} &=& \frac{|f^{\text{L}}_{\fin \ini } - f^{\text{L}}_{\ini \fin}|^2 - |f^{\text{R}}_{\fin \ini } - f^{\text{R}}_{\ini \fin}|^2}
{|f^{\text{L}}_{\fin \ini } - f^{\text{L}}_{\ini \fin}|^2 + |f^{\text{R}}_{\fin \ini } - f^{\text{R}}_{\ini \fin}|^2} \,.
\end{eqnarray}
We will not discuss the Majorana case further here since the asymmetries are similarly straightforward to obtain once coefficients of the transition dipole moment are ascertained.

At the end of this section, we comment on $CP$ violation in electrically charged neutrino decay. In this scenario, the magnitudes of the neutrino and antineutrino decay modes are modified to
\begin{eqnarray} \label{eq:amplitudes_Q}
\mathcal{M}(\nu_{\ini} \to \nu_{\fin} + \gamma_+) &=& +\sqrt{2} f^{\text{L}}_{\fin \ini } (m_\ini^2 - m_\fin^2) 
- \sqrt{2} f^Q_{\fin \ini } (m_\ini - m_\fin) \,, \nonumber\\
\mathcal{M}(\nu_{\ini} \to \nu_{\fin} + \gamma_-) &=& - \sqrt{2} f^{\text{R}}_{\fin \ini } (m_\ini^2 - m_\fin^2)
+ \sqrt{2} f^Q_{\fin \ini } (m_\ini - m_\fin) \,, \nonumber\\
\mathcal{M}(\bar{\nu}_{\ini} \to \bar{\nu}_{\fin} + \gamma_+) &=& +\sqrt{2} f^{\text{L}}_{\ini \fin} (m_\ini^2 - m_\fin^2)
- \sqrt{2} f^Q_{\ini \fin } (m_\ini - m_\fin) \,, \nonumber\\
\mathcal{M}(\bar{\nu}_{\ini} \to \bar{\nu}_{\fin} + \gamma_-) &=& -\sqrt{2} f^{\text{R}}_{\ini \fin} (m_\ini^2 - m_\fin^2)
+ \sqrt{2} f^Q_{\ini \fin } (m_\ini - m_\fin) \,,  
\end{eqnarray}
where, according to the CPT theorem, $\bar{f}^Q_{\ini \fin } = - f^Q_{\ini \fin }$ has been used. 
The modified amplitudes are equivalent to shifting coefficients $f^{\text{L}}$ and $f^{\text{R}}$ in Eqs.~\eqref{eq:amplitudes_1} and \eqref{eq:amplitudes_2} to $f^{\text{L} \prime} = f^{\text{L}} - f^Q/(m_\ini + m_\fin)$ and $f^{\text{R} \prime} = f^{\text{R}} - f^Q/(m_\ini + m_\fin)$ respectively. $CP$ asymmetries $\Delta_{CP,+}$, $\Delta_{CP,-}$, $\Delta_{CP}$ and the asymmetry between polarised photons $\Delta_{+-}$ (Dirac neutrino), as well as $\Delta_{+-}^\text{M}$ (Majorana neutrino), are obtained following the same coefficient shifts.

\section{Calculating $CP$ violation in radiative decay \label{sec:CP}}

Having provided a very general discussion on $CP$ violation and circular polarisation for neutrino radiative decay in a mass scale and model independent way in the previous section, in the following sections, we will concentrate on a simplified example where a sterile neutrino radiatively decays $\nu_s \to \nu_i + \gamma$ and show how to obtain the exact form of the $CP$ asymmetry and circular polarisation for the radiated photon. 
In this example, the initial and final state neutrinos are specified as $\nu_\ini = \nu_s$ and $\nu_\fin = \nu_i$ respectively. In this simplified case, we consider only one sterile neutrino generation and the three active neutrino generations with both $\nu_s$ and $\nu_i$ (for $i=1,2,3$) being mass eigenstates. Extensions to multiple sterile neutrino generations are straightforward, and thus, will not be discussed here. 

We will apply the above formulation in the following way. First, we estimate the size of $CP$ violation from the SM contribution alone i.e. via the charged current interaction mediated by the $W$ boson. Then, we consider the enhancement of $CP$ violation by including NP Yukawa interactions for sterile neutrinos. Such Yukawa interactions have a wide array of applications with theoretical and phenomenological utility which we will outline in the following section. Finally, we list the simplified analytical result for $CP$ violation and circular polarisation generated from the decay at the end of this section.

\subsection{The Standard Model contribution \label{sec:SM_contribution}}

It is well known that the radiative decay can happen via one-loop corrections induced by SM weak interactions with SM particles (specifically with charged lepton $\ell_\alpha$ for $\alpha = e, \mu, \tau$ and the $W$ boson) in the loop. 
The crucial operator is the charged-current interaction is
\begin{eqnarray}
\mathcal{L}_{\rm c.c.} = \sum_{\alpha=e,\mu,\tau} \; \sum_{m=1,2,3,s} \frac{g}{\sqrt{2}} U_{\alpha m} \; \bar{\ell}_\alpha \gamma^\mu  P_{\rm L} \nu_m W^{-}_{\mu} + {\rm h.c.}\,,
\end{eqnarray}
where $g$ is the EW gauge coupling constant and $U_{\alpha m}$ represent the lepton flavour mixing. Here we have $m=i,s$ (where $i=1,2,3$) representing the active light neutrino mass eigenstate $\nu_i$ and the sterile neutrino mass eigenstate $\nu_s$. 

\begin{figure}[h!]
\includegraphics[width=1\textwidth]{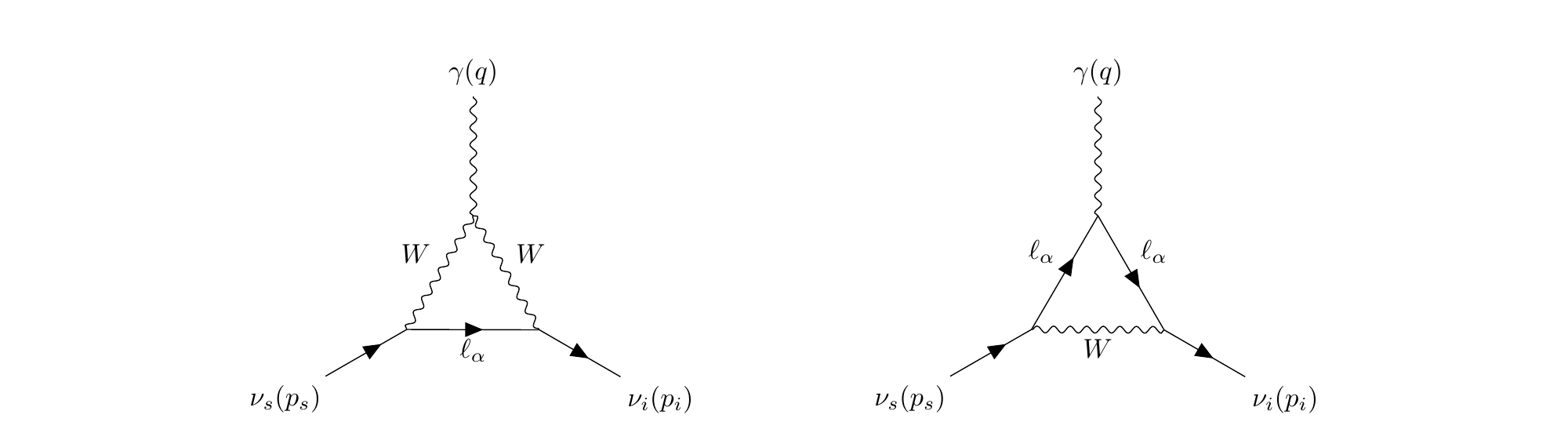}
  \caption{The Feynman diagrams for the one-loop Standard Model contributions from charged current interactions are shown above for radiative decay of a sterile neutrino. Diagrams involving unphysical Goldstone bosons and ghosts are omitted for the sake of brevity. }\label{fig:loop_SM_ints}
\end{figure}   

The one-loop Feynman diagrams for the radiative decay via the SM charged current interaction are shown in Fig.~\ref{fig:loop_SM_ints}.\footnote{In the Feynman gauge, additional diagrams involving unphysical Goldstone bosons and ghosts should also be included, note that these are not shown in the figure. In addition, the one-loop $\gamma-Z$ self-energy diagrams are essential to include to eliminate divergences in the presence of the sterile neutrino \cite{Xing:2012gd}. } 
In the limit $m_s^2/m_W^2 \ll a_\alpha \equiv m_\alpha^2/m_W^2$, where $m_\alpha$ and $m_W$ are the charged lepton and $W$ boson masses respectively, we have the result for $\Gamma^\mu_{\fin \ini}$ given as
\begin{eqnarray} \label{eq:EDM_SM}
\Gamma^\mu_{is} &=&
\frac{ieG_\text{F}\sigma^{\mu\nu}q_\nu}{4\pi^2\sqrt{2}}
\sum_{\alpha=e,\mu,\tau} U^*_{\alpha i}
U_{\alpha s} F_{\alpha} (m_s P_\text{R}+m_i P_\text{L})  \,, 
\end{eqnarray}
where $F_{\alpha}$ is a function obtained from the loop integrals and the Fermi constant is defined $\displaystyle G_{\rm F} = \frac{g^2}{4\sqrt{2} m_W^2}$. If $m_\ini$ is much smaller than the charged lepton masses, we arrive at the classic result  \cite{Pal:1981rm,Shrock:1982sc}
\begin{eqnarray}
F_\alpha =
\frac{3}{4}\left(\frac{2-a_\alpha}{1-a_\alpha}-\frac{2
a_\alpha}{(1-a_\alpha)^2}-\frac{2 a^2_\alpha\ln
a_\alpha}{(1-a_\alpha)^3}\right) \approx \frac{3}{2} - \frac{3}{4} a_\alpha\,,
\end{eqnarray} 
which is insensitive to neutrino masses. 
A more general neutrino mass-dependent result for $F_{\alpha}$ with $m_{\ini}$, $m_{\fin}$ up to the $W$ boson mass has been given in \cite{Dvornikov:2003js,Dvornikov:2004sj}. In general, for $m_{\ini}< m_W$, $F_\alpha$ is always positive, this is consistent with the optical theorem. 

From the above formulae, we obtain results for $f^{\text{L}}_{\fin \ini}$ and $f^{\text{R}}_{\fin \ini}$ given as
\begin{eqnarray}
f^{\text{L}}_{is} =
e\frac{g^2}{2} \frac{1}{16\pi^2 m_W^2}
\sum_{\alpha=e,\mu,\tau} U^*_{\alpha i}
U_{\alpha s} F_\alpha m_i  \,, \quad
f^{\text{R}}_{is} =
e\frac{g^2}{2} \frac{1}{16\pi^2 m_W^2}
\sum_{\alpha=e,\mu,\tau} U^*_{\alpha i}
U_{\alpha s} F_\alpha m_s  \,,
\end{eqnarray}
factorising the SM contribution into a coefficient part and a purely kinetic part yields
\begin{eqnarray}
f^{\text{L}}_{\fin \ini , \rm SM} = \sum_\alpha C_{\alpha} K^{\text{L}}_{\alpha} \,, \quad
f^{\text{R}}_{\fin \ini , \rm SM} = \sum_\alpha C_{\alpha} K^{\text{R}}_{\alpha}  
\end{eqnarray}
with 
\begin{eqnarray}
(C_{\alpha})_{is} &=& e \frac{g^2}{2} U_{\alpha i}^* U_{\alpha s} \,, 
\end{eqnarray}
and
\begin{eqnarray}
(K^{\text{L}}_{\alpha})_{is}  = \frac{1}{16\pi^2m_W^2} F_\alpha m_i \,, \quad
(K^{\text{R}}_{\alpha})_{is}  = \frac{1}{16\pi^2m_W^2} F_\alpha m_s \,,
\end{eqnarray}
with flavour index $\alpha=e,\mu,\tau$. 
Since $F_\alpha$ is real, both $\text{Im}(K^{\text{L}}_{\alpha} K^{\text{L}\,*}_{\beta})$ and $\text{Im}(K^{\text{R}}_{\alpha} K^{\text{R}\,*}_{\beta}
)$ vanish for any flacours $\alpha,\beta=e,\mu,\tau$. In addition, by interchanging $i \leftrightarrow s$ we notice that the one-loop SM contribution exactly satisfies $f^{\text{L}}_{\fin \ini} = \bar{f}^{\text{R}}_{\ini \fin}$ and $f^{\text{R}}_{\fin \ini} = \bar{f}^{\text{L}}_{\ini \fin}$. Therefore, there is no $CP$ violation coming from these diagrams.  

For a sterile neutrino with mass smaller than the $W$ boson mass, we comment that a non-zero $CP$ violation can in principle be obtained after considering higher-loop SM contributions.  We analyse this by applying the optical theorem once again. In order to generate an imaginary part for the kinetic loop contribution, the requirement of on-shell intermediate states has to be satisfied. 
Thus only neutrinos and photons are left in the intermediate state $c$. There are typically three cases with intermediate states given by (a)
$c = \nu_j+\gamma$,\footnote{$CP$ violation for this case has been calculated in \cite{Bell:2008fm}} (b)
$\nu_j+\nu_k+\bar{\nu}_k$, 
and (c)
$\nu_j+\alpha+\bar{\alpha}$ for $\alpha = e, \mu, \tau$. They correspond to four-, three- and two-loop diagrams respectively. 
Case (c) applies only if $m_s > 2 m_\alpha$, these contributions are in general very small. In order to obtain large $CP$ violation, additional loop contributions from NP have to be considered.

Namely, if the sterile neutrino is heavier than the $W$ boson, an imaginary part can be obtained directly from the SM one-loop diagram,  we will discuss this case in some of the following sections.

\subsection{Enhancement by new physics\label{sec:new_int}} 

In order to enhance the $CP$ violation in the radiative decay of the sterile neutrino, we include NP contributions. 
We being by introducing two new particles, one fermion $\psi$ and one scalar $\phi$ with opposite electric charges $Q$ and $-Q$ respectively. Their couplings with neutrinos and the sterile neutrino are described by the following Yukawa interaction
\begin{eqnarray} \label{eq:NP}
-\mathcal{L}_{\rm NP} \supset 
 \sum_{m=1,2,3,s} 
 \lambda_{m} \bar{\psi} \phi^* P_{\text{L}} \nu_m 
 + \lambda_{m}^* \bar{\nu}_m \phi P_{\text{R}} \psi \,,
\end{eqnarray}
where $\lambda_m$, with $m=i,s$ (for $i=1,2,3$), are complex coefficients to $\nu_i$ and $\nu_s$, which are the active and sterile neutrino mass eigenstates respectively. Here, we only included one generation of $\phi$ and $\psi$ respectively. The extension to more generations is straightforward and will be mentioned as necessary. 
Neither $\psi$ or $\phi$ are supposed to be a specific DM candidate in this work and they can annihilate with their antiparticles due to their opposite electric charges.

\begin{figure}[h!]
\includegraphics[width=1\textwidth]{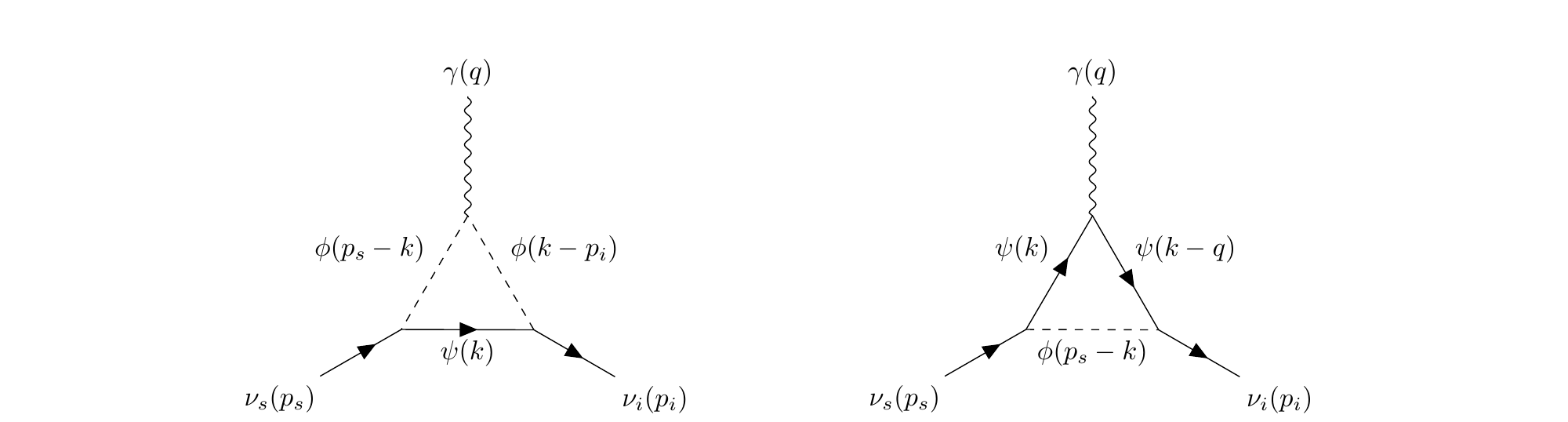}
  \caption{Feynman diagrams for the new physics one-loop contributions to the radiative decay of a sterile neutrino. We denote amplitudes for the two diagrams as $\mathcal{M}^{\rm NP}_1$ and $\mathcal{M}^{\rm NP}_2$. 
  For $\mathcal{M}^{\rm NP}_1$ we make the momenta assignments $p_1=p_s-k$, $p_2=k-p_i$ and for $\mathcal{M}^{\rm NP}_2$, we assign $k'=k-q$. 
  In both diagrams $p_s=p_i+q$. }\label{fig:loop_new_ints}
\end{figure}   

The full amplitude including the NP contribution for $\nu_s \to \nu_i + \gamma$ can then be written
\begin{eqnarray}
\mathcal{M} = \sum_\alpha \mathcal{M}^{\rm SM}_{\alpha} + \sum_{l_\text{NP}} \mathcal{M}^{\rm NP}_{l_\text{NP}} \,, 
\end{eqnarray}
where we have flavour index $\alpha = e,\mu,\tau$ and $l_\text{NP}$ represents one-loop NP contributions. Since $U(1)_Q$ is explicitly conserved and no electric charges are assigned for neutrinos at tree level, they keep free of electric charges after loop corrections are included. Thus, radiative decays are induced only via the electromagnetic transition dipole moments.
The coefficients $f^{\text{L}}_{\fin \ini }$, $f^{\text{R}}_{\fin \ini }$ and $f^{\text{L}}_{\ini \fin}$,  $f^{\text{R}}_{\ini \fin}$, including NP, are now written as 

\begin{eqnarray} \label{eq:para_BSM}
f^{\text{L}}_{\fin \ini } = \sum_\alpha C_{\alpha} K_{\alpha}^{\text{L}} + \sum_{l_\text{NP}} C_{l_\text{NP}} K_{l_\text{NP}}^{\text{L}} \,, &&  
f^{\text{R}}_{\fin \ini } = \sum_\alpha C_{\alpha} K_{\alpha}^{\text{R}} + \sum_{l_\text{NP}} C_{l_\text{NP}} K_{l_\text{NP}}^{\text{R}} \,, \nonumber\\
f^{\text{L}}_{\ini \fin} = \sum_\alpha C_{\alpha} K_{\alpha}^{\text{R}} + \sum_{l_\text{NP}} C_{l_\text{NP}} K_{l_\text{NP}}^{\text{R}} \,, &&
f^{\text{R}}_{\ini \fin} = \sum_\alpha C_{\alpha} K_{\alpha}^{\text{L}} + \sum_{l_\text{NP}} C_{l_\text{NP}} K_{l_\text{NP}}^{\text{L}} \,. 
\end{eqnarray}
From Eq.~\eqref{eq:para_BSM}, we have the necessary expressions to compute the $CP$ violation and asymmetry between the radiated photons $\gamma_+$ and $\gamma_-$. As an example, we take $\Delta_{CP,-}$ to demonstrate an explicit calculation. The definition of $\Delta_{CP,-}$ has been given in Eq.~\eqref{eq:CP_v1} where $\Delta_{CP,-} \propto |f^{\text{R}}_{\fin \ini }|^2 - |f^{\text{L}}_{\ini \fin}|^2$. With the help of the parametrisation in Eq.~\eqref{eq:para_BSM} and assuming $|K^{\text{R}}_l| = |\bar{K}^{\text{L}}_l|$ for any loop $l$, we obtain
\begin{eqnarray} \label{eq:CK_NP}
|f^{\text{R}}_{\fin \ini }|^2 - |f^{\text{L}}_{\ini \fin}|^2 &=& 
- 4 \sum_{\alpha,l_\text{NP}} {\rm Im}(C_{\alpha} C_{l_\text{NP}}^*) 
{\rm Im}(K_{\alpha}^{\text{R}} K_{l_\text{NP}}^{\text{R}\,*})  
- 2 \sum_{l_\text{NP} \neq l^{\prime}_\text{NP}} {\rm Im}(C_{l_\text{NP}} C_{l^{\prime}_\text{NP}}^*) 
{\rm Im}(K_{l_\text{NP}}^{\text{R}} K_{l^{\prime}_\text{NP}}^{\text{R}\,*}).
\end{eqnarray}

For the two NP diagrams shown in Fig.~\ref{fig:loop_new_ints}, where a photon is radiated via the interaction between scalars $\phi$ and fermions $\psi$ respectively, the amplitudes can be explicitly written as
\begin{eqnarray}
i \mathcal{M}_1^{\rm NP}(\nu_s\to\nu_i+\gamma_-)&=& - Qe  \lambda_s \lambda_i^* \int \frac{d^4 k}{(2\pi)^4}\frac{\overline{u}(p_i) P_{\text{R}} (\slashed{k}+m_\psi)(p_1-p_2)^\mu  P_{\text{L}} u(p_s)\varepsilon^*_{-,\mu}(q)}{(k^2-m_\psi^2+i\epsilon)((k-p_s)^2-m_\phi^2+i\epsilon)((k-p_i)^2-m_\phi^2+i\epsilon)} \,,\nonumber\\
i \mathcal{M}_2^{\rm NP}(\nu_s\to\nu_i+\gamma_-)&=& + Qe  \lambda_s \lambda_i^* \int \frac{d^4 k}{(2\pi)^4}\frac{\overline{u}(p_i) P_{\text{R}} (\slashed{k}'+m_\psi)\gamma^\mu(\slashed{k}+m_\psi) P_{\text{L}} u(p_s)\varepsilon^*_{-,\mu}(q)}{((k-p_s)^2-m_\phi^2+i\epsilon)(k'^2-m_\psi^2+i\epsilon)(k^2-m_\psi^2+i\epsilon)}\,.\label{eq:amplitude_no_chiral}
\end{eqnarray}
The coefficients $C_{l_\text{NP}}$ (for $l_\text{NP}=1,2$) are then simply obtained from inspection to be
\begin{eqnarray}
C_1 &=& -C_2 = -Q e \lambda_s \lambda_i^* \,.
\end{eqnarray}
In this case, ${\rm Im}(C_1 C_2^*) = 0$ and the second part of Eq.~\eqref{eq:CK_NP} vanishes. 
On the other hand the imaginary part is given by 
\begin{eqnarray}
\text{Im}(C_{\alpha} C_1^*) = - \text{Im}(C_{\alpha} C_2^*) = -\frac{Q}{2} e^2 g^2 \, \text{Im} (U_{\alpha s} U_{\alpha i}^* \lambda_i \lambda_s^* )\,.
\end{eqnarray}

We now turn to the loop contributions. $\text{Im}(K_1^{\text{R}} K_2^{\text{R}\,*}) $ does not need to be calculated since $\text{Im}(C_1 C_2^*)$ vanishes explicitly.  
Hence, the remaining term to be computed is $\text{Im}(K_{\alpha}^{\text{R}} K_{l_\text{NP}}^{\text{R}\,*}) $. 
Furthermore, since the SM contributions are always real, 
${\rm Im}(K_{\alpha}^{\text{R}} K_{l_\text{NP}}^{\text{R}\,*}) = 
- K_{\alpha}^{\text{R}} \text{Im}(K_{l_\text{NP}}^{\text{R}})$. 

In order to obtain $CP$ violation between the radiative decay $\nu_s \to \nu_i + \gamma_-$ and its $CP$ conjugate channel $\bar{\nu}_s \to \bar{\nu}_i + \gamma_+$ for a Dirac-type sterile neutrino, a non-vanishing imaginary part $\text{Im}(K_{l_\text{NP}}^{\text{R}})$ is required, this can be summarised 
\begin{eqnarray} \label{eq:K_NP}
|f^{\text{R}}_{\fin \ini }|^2 - |f^{\text{L}}_{\ini \fin}|^2 &=& 
+ 4 \sum_{\alpha,l_\text{NP}} {\rm Im}(C_{\alpha} C_{l_\text{NP}}^*) 
K_{\alpha}^{\text{R}} \text{Im}(K_{l_\text{NP}}^{\text{R}})\,.
\end{eqnarray}
Following a similar approach to determine $CP$ violation between $\nu_s \to \nu_i + \gamma_+$ and its $CP$ conjugate process $\bar{\nu}_s \to \bar{\nu}_i + \gamma_-$, we obtain 
\begin{eqnarray} 
|f^{\text{L}}_{\fin \ini }|^2 - |f^{\text{R}}_{\ini \fin}|^2 &=& 
+ 4 \sum_{\alpha,l_\text{NP}} {\rm Im}(C_{\alpha} C_{l_\text{NP}}^*) 
K_{\alpha}^{\text{L}} \text{Im}(K_{l_\text{NP}}^{\text{L}}) \,.
\end{eqnarray}

Due to the optical theorem, non-zero $\text{Im}(K_{l_\text{NP}}^{\text{L}})$ and $\text{Im}(K_{l_\text{NP}}^{\text{R}})$ can only be achieved if the sterile neutrino mass is larger than the sum of the charged scalar and the charged fermion masses, $m_s > m_\phi+m_\psi$. 
In the remainder of this section, our aim will be to compute these quantities. 
 
Here, the loop integrals for 
the relevant diagrams shown in Fig.~\ref{fig:loop_new_ints} will be calculated. Starting from the general form of the amplitude for sterile neutrino radiative decay  $\nu_s \to \nu_i + \gamma_{\pm}$ given in Eq.~\eqref{eq:decay_amplitude}, we extract the purely kinetic terms
$K^{\text{L}}_{l_\text{NP}}$ and $K^{\text{R}}_{l_\text{NP}}$ for $l_\text{NP}=1,2$ as
\footnote{Here, $K^{\text{L}}_{l_\text{NP}}$ and $K^{\text{R}}_{l_\text{NP}}$ represent $(K^{\text{L}}_{l_\text{NP}})_{is}$ and $(K^{\text{R}}_{l_\text{NP}})_{is}$, respectively. Exchanging $i$ with $s$, we obtain $(K^{\text{L}}_{l_\text{NP}})_{si} = (K^{\text{R}}_{l_\text{NP}})_{is}$ and $(K^{\text{R}}_{l_\text{NP}})_{si} = 0$, this is compatible with our previous statement that $(K^{\text{L}}_l)_{\ini \fin} = (K^{\text{R}}_l)_{\fin \ini}$ and $(K^{\text{R}}_l)_{\ini \fin} = (K^{\text{L}}_l)_{\fin \ini}$. }
\begin{eqnarray}\label{eq:form_f1}
K^{\text{L}}_1 = \frac{m_i}{16\pi^2} \int_0^1\dx\dy\dz\frac{\delta(x+y+z-1)\, z}{\Delta_{\phi\psi}(x,y,z)}\,,\;\; && 
K^{\text{R}}_1 = \frac{m_s}{16\pi^2} \int_0^1\dx\dy\dz\frac{\delta(x+y+z-1)\, y}{\Delta_{\phi\psi}(x,y,z)} \,,\nonumber\\
K^{\text{L}}_2 = \frac{m_i}{16\pi^2} \int_0^1\dx\dy\dz\frac{\delta(x+y+z-1)\,xz}{\Delta_{\psi\phi}(x,y,z)}\,, && 
K^{\text{R}}_2 = \frac{m_s}{16\pi^2} \int_0^1\dx\dy\dz\frac{\delta(x+y+z-1)\,xy}{\Delta_{\psi\phi}(x,y,z)}\,,
\end{eqnarray}
where 
\begin{eqnarray} \label{eq:Deltaxyz}
\Delta_{\phi\psi}(x,y,z) &=& m_\phi^2(1-x)+x m_\psi^2-x (y m_s^2 + z m_i^2) \nonumber\\
\Delta_{\psi\phi}(x,y,z) &=& m_\psi^2(1-x)+x m_\phi^2-x (y m_s^2 + z m_i^2) \,. 
\end{eqnarray}  
The above results are obtained without any approximations. In order to derive further simplified analytical formulae, we consider the large mass hierarchy between $\nu_s$ and $\nu_i$ where $m_i \ll m_s$, and may therefore take the limit $m_i \to 0$. In this case, $K^{\text{L}}_{l_\text{NP}}=0$ and after integrating over Feynman parameters $z$ and $x$, $K^{\text{R}}_{l_\text{NP}}$ can be written as
\begin{eqnarray}\label{eq:Kzx}
K^{\text{R}}_1 &=& \frac{m_s}{16\pi^2} \int_0^1\dy\frac{ y }{m_s^2 y-m_\psi^2+m_\phi^2} \log \left(\frac{\Delta_{\phi\psi}(y)}{m_\phi^2}\right) \,, \nonumber\\
K^{\text{R}}_2 &=& \frac{m_s}{16\pi^2} \left[ \int_0^1\dy \frac{ - m_\psi^2 y}{(m_s^2 y+m_\psi^2-m_\phi^2 )^2}
\log \left(\frac{\Delta_{\psi\phi}(y)}{m_\psi^2}\right)
+ \int_0^1\dy\frac{y (y-1)}{m_s^2 y+m_\psi^2-m_\phi^2} \right]\,,
\end{eqnarray}
where 
\begin{eqnarray}
\Delta_{\phi\psi}(y) &=& y \left(m_s^2 (y-1)+m_\phi^2\right)-m_\psi^2 (y-1) \,, \nonumber\\
\Delta_{\psi\phi}(y) &=& y \left(m_s^2 (y-1)+m_\psi^2\right)-m_\phi^2 (y-1) \,.
\end{eqnarray} 

$K^{\text{R}}_{l_\text{NP}}$ may have both real parts and imaginary parts. 
The real part $\text{Re}(K^{\text{R}}_{l_\text{NP}})$ is directly obtained by replacing $\Delta_{\phi\psi}$ and $\Delta_{\psi\phi}$ with there absolute values, therefore simple analytical expressions for $\text{Re}(K^{\text{R}}_{l_\text{NP}})$ are difficult to obtain. However, in the hierarchical case $m_s\gg m_\phi, m_\psi$ approximate analytical expressions can be derived by expanding in powers of $m_\phi^2/m_s^2$ and $m_\psi^2/m_s^2$. Specifically,  the leading-order results are given by 
\begin{eqnarray}
\text{Re}(K^{\text{R}}_1) &\approx& \frac{1}{16\pi^2 m_s} \left[ \log \left( \frac{m_s^2}{m_\phi^2} \right) - 2 \right] \,, \nonumber\\
\text{Re}(K^{\text{R}}_2) &\approx& \frac{1}{16\pi^2 m_s} \times \frac{-1}{2} \,. 
\end{eqnarray}
Since we are chiefly interested in the $CP$ violating component, we will focus on how to obtain and simplify the imaginary parts of $K^{\text{R}}_{l_\text{NP}}$. 

Since  $m_{\phi}^2, m_{\psi}^2 \geq 0$,  the imaginary and thus $CP$ violating component in  Eq.~\eqref{eq:Kzx} factorises when the argument of the logarithm is negative, by inspection we can see this occurs when
\begin{eqnarray}\label{eq:cond_CP}
\Delta_{\phi\psi}(y) &<& 0 \,, \nonumber\\
\Delta_{\psi\phi}(y) &<& 0 \,.
\end{eqnarray} 
Solutions at the boundaries of the $CP$ violation conditions $\Delta_{\phi\psi}(y)=0$ and $\Delta_{\psi\phi}(y)=0$ are $y_{1,2}(m_{\phi},m_{\psi})$ and $y_{1,2}(m_{\psi},m_{\phi})$ respectively. Therefore the conditions in Eq. (\ref{eq:cond_CP}) in terms of $y$ are fulfilled when $y_1(m_\phi,m_\psi)\leq y\leq y_2(m_\phi,m_\psi)$ and $y_1(m_\psi,m_\phi)\leq y\leq y_2(m_\psi,m_\phi)$ for the two diagrams respectively, where
\begin{eqnarray}
y_{1,2}(m_\phi,m_\psi) = \frac{1}{2} + \frac{m_\psi^2  - m_\phi^2 \mp \mu^2}{2 m_s^2} \,,\nonumber\\
y_{1,2}(m_\psi,m_\phi) = \frac{1}{2} + \frac{m_\phi^2  - m_\psi^2 \mp \mu^2}{2 m_s^2} \,,
\end{eqnarray}
and $\mu^2$ is defined as
\begin{eqnarray}
\mu^2 = \sqrt{m_s^4 + m_\phi^4 + m_\psi^4 - 2 m_s^2 m_\phi^2 - 2 m_s^2 m_\psi^2 - 2 m_\phi^2 m_\psi^2} \,. 
\end{eqnarray}
It should be noted that in both cases $0< y_1<y_2<1$ is necessarily satisfied.

Hence, the imaginary component of Eq.~\eqref{eq:Kzx} can now be written according to  the complex logarithm definition as 
\begin{eqnarray}\label{eq:im_ky}
\mathrm{Im}(K^{\text{R}}_1) &=& \frac{m_s}{16\pi^2} \times \pi \int_{y_1(m_\phi,m_\psi)}^{ y_2(m_\phi,m_\psi)}\dy\frac{y }{m_s^2 y-m_\psi^2+m_\phi^2} \,, \nonumber\\
\mathrm{Im}(K^{\text{R}}_2) &=& \frac{m_s}{16\pi^2} \times \pi  \int_{y_1(m_\psi,m_\phi)}^{y_2(m_\psi,m_\phi)}\dy\frac{-m_\psi^2 y}{(m_s^2 y+m_\psi^2-m_\phi^2)^2} \,.
\end{eqnarray}
Finally, integrating over the final Feynman parameter $y$ leads to 
\begin{eqnarray}\label{eq:im_k}
\mathrm{Im}(K^{\text{R}}_1)&=& \frac{m_s}{16\pi^2} \frac{ - \pi}{m_s^2} \left[\frac{\mu^2}{m_s^2}+\frac{m_\phi^2-m_\psi^2}{m_s^2} \log \left(\frac{m_s^2+m_\phi^2-m_\psi^2-\mu^2}{m_s^2+m_\phi^2-m_\psi^2+\mu^2}\right)\right] \,,\nonumber\\
\mathrm{Im}(K^{\text{R}}_2)&=& \frac{m_s}{16\pi^2} \frac{+ \pi}{m_s^2}\left[\frac{\mu^2(m_\psi^2-m_\phi^2)}{m_s^4} + \frac{m_\psi^2}{m_s^2} \log \left(\frac{m_s^2+m_\psi^2-m_\phi^2-\mu^2}{m_s^2+m_\psi^2-m_\phi^2+\mu^2}\right)\right] \,.
\end{eqnarray}
The requirement $m_s > m_\phi + m_\psi$ leads to a positive $\mu^2$. In the mass-degenerate limit $m_s = m_\phi + m_\psi$, $\mu^2 = 0$ and after some simplifications, it can be shown for this case that $\mathrm{Im}(K^{\text{R}}_1) = \mathrm{Im}(K^{\text{R}}_2) = 0$. In the massless limit $m_\phi, m_\psi \to 0$, these imaginary parts are approximately given by $\mathrm{Im}(K^{\text{R}}_1) \to -1/(16 \pi m_s)$ and $\mathrm{Im}(K^{\text{R}}_2) \to 0$.

Since we need to compute $\Delta_{CP,-}$ to calculate $CP$ violation, we apply Eq.~\eqref{eq:K_NP}, which in this example can be written explicitly as $|f^{\text{R}}_{\fin \ini }|^2 - |f^{\text{L}}_{\ini \fin}|^2 = 
+ 4 \sum_{\alpha} {\rm Im}(C_{\alpha} C_1^*) 
K_{\alpha}^{\text{R}} [ \text{Im}(K_1^{\text{R}} - K_2^{\text{R}})]$, therefore we obtain
\begin{eqnarray}|f^{\text{R}}_{\fin \ini }|^2 - |f^{\text{L}}_{\ini \fin}|^2 = 
\frac{2 \pi Q e^2 g^2}{(16\pi^2)^2 m_W^2} \sum_{\alpha} \text{Im} (U_{\alpha s} U_{\alpha i}^* \lambda_i \lambda_s^* ) \, F_\alpha \, I_{\phi\psi} \,.
\end{eqnarray}
For $\Delta_{CP,+}$, $|f^{\text{R}}_{\fin \ini }|^2 - |f^{\text{L}}_{\ini \fin}|^2$ is obtained by multiplying by a factor $m_i^2/m_s^2$ which is strongly suppressed by the light active neutrino mass. 

Here, we have defined $I_{\phi\psi}$, an order one normalised parameter which is defined via
$\displaystyle \text{Im}(K_2^{\text{R}} - K_1^{\text{R}}) = \frac{m_s}{16\pi^2} \frac{ \pi}{m_s^2} I_{\phi\psi}$ and explicitly given by 
\begin{eqnarray} 
I_{\phi\psi} = 
\frac{\mu^2(m_s^2 +m_\psi^2-m_\phi^2)}{m_s^4} + 
\frac{m_\phi^2-m_\psi^2}{m_s^2} \log \left(\frac{m_s^2+m_\phi^2-m_\psi^2-\mu^2}{m_s^2+m_\phi^2-m_\psi^2+\mu^2}\right)&& \nonumber\\
+ \frac{m_\psi^2}{m_s^2} \log \left(\frac{m_s^2+m_\psi^2-m_\phi^2-\mu^2}{m_s^2+m_\psi^2-m_\phi^2+\mu^2}\right)&.& 
\end{eqnarray}
See Appendix \ref{sec:imaginary_detail} for more details regarding the calculation of the imaginary part of the loop diagrams.

In this example, we may safely ignore the $f^{\text{L}}_{\fin \ini }$ and $f^{\text{R}}_{\ini \fin}$ terms since $f^{\text{L}}_{\fin \ini } \sim f^{\text{R}}_{\ini \fin} \sim \frac{m_i}{m_s} f^{\text{L}}_{\ini \fin} \sim \frac{m_i}{m_s} f^{\text{R}}_{\fin \ini }$, thus the asymmetries, defined in Eqs.~\eqref{eq:CP_v1} and \eqref{eq:CP_v2} are approximately given by
\begin{eqnarray} \label{eq:CP_D_v3}
 - \Delta_{CP,-} \approx - \Delta_{CP} \approx \Delta_{+-}\approx \frac{|f^{\text{L}}_{\ini \fin}|^2 - |f^{\text{R}}_{\fin \ini }|^2}
{|f^{\text{L}}_{\ini \fin}|^2 + |f^{\text{R}}_{\fin \ini }|^2} \,
\end{eqnarray}
and $\Delta_{CP,+}$ is negligibly small. This result works for the Dirac neutrino case. 
In the Majorana neutrino case, from Eq.~\eqref{eq:CP_M_v2}, it is straightforward to apply a similar procedure and obtain
\begin{eqnarray} \label{eq:CP_M_v3}
\Delta_{CP,+}^\text{M} = - \Delta_{CP,-}^\text{M} = \Delta_{+-}^\text{M} \approx \frac{|f^{\text{L}}_{\ini \fin}|^2 - |f^{\text{R}}_{\fin \ini }|^2}
{|f^{\text{L}}_{\ini \fin}|^2 + |f^{\text{R}}_{\fin \ini }|^2} 
\end{eqnarray}
and $\Delta_{CP} = 0$. Regardless of whether the neutrinos are Dirac or Majorana particles $\Delta_{CP,-} \approx -\Delta_{+-}$ is satisfied. This is true in general if $f^{\text{L}}_{\fin \ini }, f^{\text{R}}_{\ini \fin} \ll f^{\text{L}}_{\ini \fin}, f^{\text{R}}_{\fin \ini }$.

\section{Phenomenological applications of the formulation\label{sec:applications}}

We are now ready to discuss possible phenomenological implications of this suggested sterile neutrino model which has $CP$ violation generated at one-loop level for radiative decays. The formulation based on the simplified example above has a wide array of possible applications. 
One direct application is the study of $CP$ violation in keV neutrino DM radiative decay. We can also apply it to the general type-I seesaw mechanism where right-handed neutrinos are much heavier than the electroweak scale in order to recover light active neutrino masses. It is also of interest to consider its application for heavy neutrino DM motivated by the IceCube data.

\subsection{keV sterile neutrino dark matter }
The keV-scale sterile neutrino has been discussed extensively as a DM candidate (for example models, see \cite{Li:2010vy, Araki:2011zg, Abazajian:2014gza, Vincent:2014rja, Harada:2014lma}). 
Following the discussion in Section \ref{sec:SM_contribution}, it is clear that the SM contribution at one-loop level cannot generate $CP$ violation in keV neutrino radiative decay and a non-zero $CP$ asymmetry can only be obtained at four-loop level. Therefore, we consider Yukawa interactions as shown  in Eq.~\eqref{eq:NP}. 

We give a brief discussion on constraints to the sterile neutrino $\nu_s$ and the new charged particles $\phi$ and $\psi$. 
Since $\nu_s$ is assumed to be a DM candidate, the decay channel $\nu_s \to \phi \psi$ introduced by the new interaction with  $\phi$ and $\psi$ must be controlled. The width of this channel is around 
\begin{eqnarray}
\Gamma_{\rm NP} = c_\nu \frac{|\lambda_s|^2}{8\pi} m_s \,,
\end{eqnarray} 
where $c_\nu = 1$ for a Dirac neutrino and $c_\nu = 2$ for a Majorana neutrino.  
We require the width to be at least as small as the decay width of the SM $\Gamma_{\rm SM}$. We approximate $\Gamma_{\rm SM}$ to the width of the dominant channels $\nu_s \to \nu_i \nu_j \bar{\nu}_i$ for any active neutrinos $\nu_i$ and $\nu_j$ \cite{Pal:1981rm,Barger:1995ty,Li:2010vy} namely 
\begin{eqnarray}
\Gamma_{\rm SM} \approx c_\nu \frac{G_F^2 m_s^5}{192\pi^3} 
\sum_{i=1,2,3}|(U^\dag U)_{is}|^2 \,,
\end{eqnarray}
where $(U^\dag U)_{is} = \sum_{\alpha=e,\mu,\tau} U_{\alpha i}^* U_{\alpha s}$. 
By introducing a parameter $\eta$ representing the ratio of the two decay widths 
$\eta = \Gamma_{\rm NP} / \Gamma_{\rm SM}$, we can express $|\lambda_s|$ by $\eta$ as 
$|\lambda_s| \approx \frac{1}{2\sqrt{6} \pi} \sqrt{\eta}\, G_F m_s^2  \sqrt{\sum_{i}|(U^\dag U)_{is}|^2}$, namely, an extremely small value for $\lambda_s$ is required.\footnote{Note that $G_F m_s^2 \sim 10^{-16}$ for keV sterile neutrino DM.}
The charged particles $\phi$ and $\psi$ as in our previous formulation are assumed to be lighter than the sterile neutrino. Thus, they have to be at most millicharged to avoid significant modification to the precisely measured QED interactions at low energy. The Lamb shift imposes an upper bound for the millicharge $Q \lesssim 10^{-4} e$ \cite{Gluck:2007ia}, which is valid for a scalar or fermion with a mass less than 1 keV.

Considering these bounds, we can roughly estimate the size of $CP$ violation of $\nu_s$ radiative decay. 
We also recall that the SM decay channel dominates the DM radiative decay while $\eta <1$.

In this case, we can approximate both $f^{\text{R}}_{\fin \ini }$ and $f^{\text{L}}_{\ini \fin}$ in the denominator by 
$f^{\text{R}}_{\fin \ini, \text{SM}}$  and it then follows that 
\begin{eqnarray}
\Delta_{CP,-} \approx \Delta_{CP} \approx - \Delta_{+-}\approx
\frac{|f^{\text{R}}_{\fin \ini }|^2 - |f^{\text{L}}_{\ini \fin}|^2}{2 |f^{\text{R}}_{\fin \ini, \text{SM}}|^2} \,. 
\end{eqnarray}
Therefore, we obtain the analytical result of the $CP$ asymmmetry as 
\begin{eqnarray}
\Delta_{CP,-} &\approx& 
\frac{8 \pi}{3} \frac{Q m_W^2}{g^2 m_s^2} 
\frac{\text{Im} ( \lambda_i (U^\dag U)_{is} \lambda_s^* )}{(U^\dag U)_{is}} 
I_{\phi\psi} \nonumber\\
&\approx& 
\frac{\sqrt{\eta}  Q}{6\sqrt{3}} |\lambda_i| I_{\phi\psi} \sin\delta_{is}
\frac{\sqrt{|(U^\dag U)_{1s}|^2+|(U^\dag U)_{2s}|^2+|(U^\dag U)_{3s}|^2}}
{(U^\dag U)_{is}}
\end{eqnarray}
where we have made the approximations $F_\alpha \approx 3/2$ since $m_\alpha \ll m_W$, and denoted the phase of $\lambda_i (U^\dag U)_{is} \lambda_s^* $ as $\delta_{is}$. In the limits $m_s \gg m_\phi, m_\psi$, we have $I_{\phi\psi} \approx 1$, and thus arrive at $\Delta_{CP,-} \sim 10^{-1} \sqrt{\eta}  Q |\lambda_i|$, which is small due to the suppression by the millicharge $Q$. Enhancement can be achieved by considering a different parameter space. For example, by assuming $m_s, m_\psi \gg m_\phi$, we have $I_{\phi\psi} \approx \frac{m_\psi^2}{m_s^2} \log\frac{(m_s^2-m_\psi^2)^2}{m_s^2 m_\phi^2}$, and thus the enhancement by an order of magnitude is easily obtained from $I_{\phi\psi}$. By assuming a typical value of the millicharge $Q \sim 10^{-4} e$, the coupling $\lambda_i \sim 10^{-1}$ and $\eta \sim 1$, we arrive at $\Delta_{CP,-}\sim 10^{-5}$. Other enhancements could be realised by considering the hierarchical mixing of the sterile neutrino with different active neutrinos. 

\subsection{Seesaw mechanism and leptogenesis }
Our discussion thus far can also be generalised to the case of very heavy neutrinos. Heavy neutrinos with masses much higher than the EW scale are introduced in the seesaw mechanism to explain the tiny observed active neutrino masses. The heavy neutrinos are assumed to be Majorana particles in the mechanism. 
These particles, as originally proposed in \cite{Fukugita:1986hr}, provide a class of scenarios where matter-antimatter asymmetry of the Universe is generated by the decays of heavy neutrinos by a process termed leptogenesis. 

Yukawa interactions involving heavy neutrinos provide the necessary source of $CP$ violation between the decay $N_I \to L_\alpha H$ and its $CP$ conjugate $N_I \to \bar{L}_\alpha H^\dag$ in leptogenesis. 
We address the fact that these interactions can also generate $CP$ violation between the radiative decay $N_I \to N_J \gamma_+$ and its $CP$ conjugate process $N_I \to N_J \gamma_-$.\footnote{Neutrinos are Majorana particles in the seesaw mechanism framework.} The $CP$ asymmetry can be simply estimated with the help of the analytical result obtained in the last subsection. 
In order to achieve this, we first present the Yukawa interactions in the form 
\begin{eqnarray} \label{eq:LHN}
- \mathcal{L}_Y  \supset  \sum_{\alpha, I} \lambda_{\alpha I} \bar{L}_\alpha \tilde{H} P_\text{R} N_I + \lambda_{\alpha I}^* \bar{N}_I \tilde{H}^\dag P_\text{L} L_\alpha
= \sum_{\alpha, I} \lambda_{\alpha I} \bar{N}_I^c \tilde{H}^T P_\text{R} L_\alpha^c + \lambda_{\alpha I}^* \bar{L}_\alpha^c \tilde{H}^* P_\text{L} N_I^c \,,
\end{eqnarray} 
where $\tilde{H} = i\sigma_2 H^*$.  
Since we consider right-handed neutrinos to be much heavier than the $W$ boson mass, the Goldstone-boson equivalence theorem can be applied. 
The main contributions to $N_I \to N_J \gamma$ are those loops involving charged leptons $\ell_\alpha$ and the Goldstone boson $H^+$. Therefore, we can simply apply the formulation in Section \ref{sec:new_int} by replacing masses $m_\psi$ and $m_\phi$ with $m_\alpha$ and $m_W$ respectively. Here, it is necessary to keep the charged lepton masses as we will see later that it is essential to generate $CP$ asymmetry. 
In this case, $f^{\text{L}}_{\fin \ini}$ and $f^{\text{R}}_{\fin \ini}$ are approximatively given by 
$f^{\text{L}}_{J I} \approx \sum_\alpha C_{\alpha} K^{\text{L}}_\alpha$
and
$f^{\text{R}}_{J I} \approx \sum_\alpha C_{\alpha} K^{\text{R}}_\alpha$ with $ C_{\alpha}$ and $K^{\text{L}, \text{R}}_\alpha$ given by
\begin{eqnarray}
&& C_{\alpha} = - e \lambda_{\alpha I} \lambda_{\alpha J}^*\,,\nonumber\\
&& K^{\text{L}}_\alpha = K^{\text{L}}_{1,\alpha} - K^{\text{L}}_{2,\alpha}\,,\quad
K^{\text{R}}_\alpha = K^{\text{R}}_{1,\alpha} - K^{\text{R}}_{2,\alpha}
\end{eqnarray} 
with $K^{\text{R}}_{1,\alpha}$ and $K^{\text{R}}_{2,\alpha}$ given by $K^{\text{R}}_{1}$ and $K^{\text{R}}_{2}$ in Eq.~\eqref{eq:form_f1} with masses $m_s$, $m_i$, $m_\phi$, $m_\psi$ replaced by $M_I$, $M_J$, $m_\alpha$ and $m_W$ respectively. Assuming right handed neutrino masses $M_I \gg M_J$, we can safely ignore the $K^{\text{L}}_{\alpha}$ contribution and arrive at the approximation of $CP$ asymmetry shown in Eq.~\eqref{eq:CP_M_v3}. 

The $CP$ violation requires both non-zero values for $\text{Im}(C_\alpha C_\beta^*)$ and $\text{Im}(K^{\text{R}}_\alpha K^{\text{R}\,*}_\beta)$. The former term given by $\text{Im}(C_\alpha C_\beta^*) = e^2 \text{Im} ( \lambda_{\alpha I} \lambda_{\alpha J}^* \lambda_{\beta J} \lambda_{\beta I}^*)$ is usually non-zero based on the complex Yukawa couplings which are necessary for leptogenesis. For the latter term, without considering the difference between charged lepton masses $K^{\text{R}}_\alpha = K^{\text{R}}_\beta$ and $\text{Im}(K^{\text{R}}_\alpha K^{\text{R}\,*}_\beta)=0$ holds explicitly. Taking charged lepton masses into account and considering the hierarchy $m_\alpha \ll m_W \ll M_I$, we obtain the leading contribution (c.f. Eq.~\eqref{eq:Kzx} and Eq.~ \eqref{eq:im_k})
\begin{eqnarray} 
\text{Im}(K^{\text{R}}_\alpha K^{\text{R}\,*}_\beta) &\approx& \frac{-\pi}{(16\pi^2 M_I)^2}  
\log \left( \frac{m_W^2}{M_I^2} \right)
\left[ \frac{m_\alpha^2}{M_I^2} \log \left( \frac{m_\alpha^2}{M_I^2} \right) 
- \frac{m_\beta^2}{M_I^2} \log \left( \frac{m_\beta^2}{M_I^2} \right) \right] \,.
\end{eqnarray} 
Eventually, we arrive at the CP asymmetry as
\begin{eqnarray} 
\Delta_{CP,-} &\approx& 
\frac{- \pi e^2 }{| [\lambda^\dag \lambda]_{IJ}|^2} ~
\text{Im}( \lambda_{\tau I} [\lambda^\dag \lambda]_{IJ} \lambda_{\tau J}^*) ~
\frac{m_\tau^2}{M_I^2} \log \left( \frac{m_\tau^2}{M_I^2} \right) 
\left/ \log \left( \frac{m_W^2}{M_I^2} \right) \right.,
\end{eqnarray}
where for charged leptons, only the dominant $\tau$ mass has been considered. This formula takes a similar structure as the $CP$ asymmetry of the $N\to L_\tau H$ decay in thermal leptogenesis (see e.g., in \cite{Davidson:2008bu}), namely, the coefficient combination, $\text{Im}( \lambda_{\tau I} [\lambda^\dag \lambda]_{IJ} \lambda_{\tau J}^*)$. The difference is that, while the asymmetry in thermal leptogenesis is suppressed by a loop factor,\footnote{The leading order contribution of the $N\to L_\tau H$ decay is at tree level and the $CP$ violation appears at one-loop level. } the asymmetry here is not, but rather strongly suppressed by the mass hierarchy $m_\tau^2 / M_I^2$. 

Furthermore, we comment that the $CP$ violation for heavy neutrino radiative decays are very hard to observe since the only way to access this quantity is to measure the circular polarisation of photons radiated from the decay. This is not possible to measure currently due to the very small size of $\Delta_{+-}$. What presents an even larger challenge is that these processes happen in the very early stages of the evolution of the Universe. Thus, even if there is a large fraction of polarised photons produced, the asymmetry will be washed out by ubiquitous Compton scattering processes \cite{Boehm:2019lvx}. 

A possible way to enhance the $CP$ asymmetry may be by considering a low-energy seesaw mechanism. For example, in the GeV sterile neutrino seesaw, there is no severe mass suppression between right-handed neutrino masses and the $\tau$ lepton mass to significantly reduce the $CP$ asymmetry. Neutrinos at such a scale can explain baryon asymmetry based on a different leptogenesis mechanism, specifically the Akhmedov-Rubakov-Smirnov mechanism \cite{Akhmedov:1998qx}. Another advantage is that these neutrinos can be tested at the SHiP experiment \cite{Anelli:2015pba}. The disadvantage is that since the neutrino mass is lower than the $W$ boson mass, $CP$ violation of the radiative decay cannot be generated at one-loop, but rather at two-loop level. Thus, a more complicated calculation is required for this case.  

\subsection{Heavy dark matter and IceCube}

Very heavy neutrinos could also be DM candidates. In fact, a heavy neutrino DM $N_{\rm DM}$ with mass around $10^2$~TeV -- PeV scale as a DM candidate \cite{Chianese:2016opp,Aartsen:2018mxl} is motivated by the  
high energy neutrino component in excess of the well-known atmospheric events \cite{Aartsen:2013bka,Aartsen:2015rwa} by the IceCube experiment (see \cite{Stettner:2019tok,Chianese:2019kyl} for recent progresses and \cite{Sui:2018bbh,Chianese:2018ijk} for analysis combining with other experimental data). Examples of typical heavy neutrino DM models explaining these observations have been shown in \cite{DiBari:2016guw,Fiorentin:2016avj,Chianese:2016smc,Hiroshima:2017hmy,DiBari:2019zcc}. 
At low energy, they may induce very weak effective Yukawa interactions between the DM neutrino with other fermions. 

Since radiative decay of a DM candidate can proceed very slowly until the present day, the washout by Compton scattering in the early stage of Universe can be avoided. 
Given a sufficiently small Yukawa coupling $\lambda_{\alpha\text{-DM} }\bar{L}_\alpha \tilde{H} N_{\rm DM}$,\footnote{This Yukawa coupling may be effectively induced. For example, in the Higgs induced RHiNo DM model \cite{DiBari:2016guw,DiBari:2019zcc,Anisimov:2008gg}, it is the dimension-five operator $\displaystyle \frac{1}{\Lambda} \bar{N}_I^c N_{\rm DM} H^\dag H$ with the thermal effect enhancing the mixing between DM with source neutrino $N_I$ which eventually enhances the DM production. This operator, together with the Yukawa coupling Eq.~\eqref{eq:LHN} induces a very weak Yukawa coupling with coefficient $\lambda_{\alpha\text{-DM}} \sim y_{\alpha I} \frac{v_H M_I}{\Lambda M_{\rm DM}}$ in the limit $M_{\rm DM} \gg M_I$ where $v_H$ is the Higgs VEV. } 
we may easily estimate the size of $CP$ asymmetry in the DM radiative decay. The tree-level decay to $\nu Z$ is induced and is one of the main decay channels being tested at IceCube. On the other hand, this coupling also induces the radiative decay $ N_{\rm DM} \to \nu \gamma$ which may result in $CP$ violation. The $CP$ violation, as discussed in the last subsection, would be suppressed by the ratio $m_\tau^2 / M_{\rm DM}^2 \lesssim 10^{-6}$.

\section{Conclusion\label{sec:conclusion}} 

In this work, we built a general framework for $CP$ violation in neutrino radiative decays. $CP$ violation in such processes produces an asymmetry between the circularly polarised radiated photons and provides an important source of net circular polarisation that can be observed in particle and astroparticle physics experiments. 

The formulation between $CP$ violation in neutrino radiative decays and the neutrino electromagnetic dipole moment at the form factor level is developed for both Dirac and Majorana neutrinos. We observed the model-independent connection between the decays and photon circular polarisation produced by these processes and concluded that $CP$ violation directly determines the circular polarisation.  
Specifically in the Majorana neutrino case, the $CP$ asymmetry is identical to the asymmetry of photon polarisations up to an overall sign difference. The contribution of a non-zero electric charge to neutrino decays is also discussed for completeness.

We then discussed how to generate non-vanishing $CP$ violation through a generic new physics Yukawa interaction extension consisting of electrically charged scalar and fermion states. Without introducing any source of electric charge for the neutrinos, these particles can decay only via the electromagnetic transition dipole moment. The explicit analytical result of $CP$ violation for this model was derived and presented. 
This fundamental result is applicable when determining circular polarisation for both Dirac and Majorana fermions and can be exported for use in any models that generate radiative decays of this type. 

Finally, we included some brief discussion pertaining to the phenomenological implications of neutrinos at various mass scales. 
Firstly, the fomalism was applied to keV sterile neutrinos which are popular DM candidates and found $CP$ violation and circular polarisation of the resulting radiated $X$-ray. 
We also considered the implications for much heavier sterile neutrinos of scale $\gtrsim 1$TeV which are required for the seesaw mechanism and leptogenesis. We argue that the $CP$ source in the Yukawa coupling, which is essential for leptogenesis, can trigger $CP$ violation for heavy neutrino radiative decays. The case of weakly interacting sterile neutrinos at a mass comparable to the electroweak scale is also interesting as it could produce exotic collider signatures as well as circular polarisation. We plan to compute the $CP$ violation from such a process in future work.
We also discussed the circular polarisation of $\gamma$-rays released from the radiative decay of the PeV scale dark matter motivated by IceCube data, however the size of this effect is too small to observe at current experimental sensitivities. 

\section*{Acknowledgements} 

We would like to thank C\'eline B\oe hm for suggesting investigation of circular polarisation produced by sterile neutrinos and initiating our collaboration.
We are also grateful to S. Arunasalam, M. Chianese, Y.-F. Li, A. Titov and S. Zhou for their useful discussions. SB is supported by the Australian Research Council (ARC).
MRQ is supported by Consejo Nacional de Ciencia y Tecnologia, Mexico (CONACyT) under grant 440771. YLZ acknowledges the STFC Consolidated Grant ST/L000296/1 and the European Union's Horizon 2020 Research and Innovation programme under Marie Sk\l{}odowska-Curie grant agreements Elusives ITN No.\ 674896 and InvisiblesPlus RISE No.\ 690575.

\appendix

\section{Polarisation-dependent amplitudes \label{sec:rest_frame}}

We may derive the amplitudes of neutrino and antineutrino radiative decays specifying the photon polarisation in the final state, $\mathcal{M}(\nu_\ini \to \nu_\fin + \gamma_\pm)$ and $\mathcal{M}(\bar{\nu}_\ini \to \bar{\nu}_\fin + \gamma_\pm)$. 

We apply the chiral representation, where the $\gamma$ matrices are given by
\begin{eqnarray}
\gamma^\mu =  \begin{pmatrix}
0 & \sigma^\mu  \\
\bar{\sigma}^\mu & 0
\end{pmatrix} \,, \quad
\sigma^{\mu\nu} = \frac{i}{2} [\gamma^\mu,\gamma^\nu] \,,\quad
\gamma_5 \equiv i\gamma^0\gamma^1\gamma^2\gamma^3 =  \begin{pmatrix}
-\mathbf{1} & 0  \\
0 & \mathbf{1}
\end{pmatrix} \,, \quad 
P_{\text{L,R}} = \frac{1\mp \gamma_5}{2},
\end{eqnarray}
and $\sigma^\mu=(\mathbf{1},\sigma^1, \sigma^2, \sigma^3)$ and $\bar{\sigma}^\mu=(\mathbf{1},-\sigma^1, -\sigma^2, -\sigma^3)$ and $\sigma^i$ are Pauli matrices. 
Given momentum $p=(p_0,\vec{p})$, the normalised particle and antiparticle Dirac spinors are represented by 
\begin{eqnarray}
u_S(p) = \begin{pmatrix}
\sqrt{p \cdot \sigma}~ \xi_S  \\
\sqrt{p \cdot \bar{\sigma}}~ \xi_S
\end{pmatrix} \,, \quad
v_S(p) = \begin{pmatrix}
\sqrt{p \cdot \sigma}~ \eta_S  \\
\sqrt{-p \cdot \bar{\sigma}}~ \eta_S
\end{pmatrix} \,,
\end{eqnarray}
where $\xi_S$ and $\eta_S$ are two-component spinors normalised to unity. Here, we include the polarisation index $S$ for two independent spinors.

\begin{figure}[h!]
\centering
\includegraphics[width=.7\textwidth]{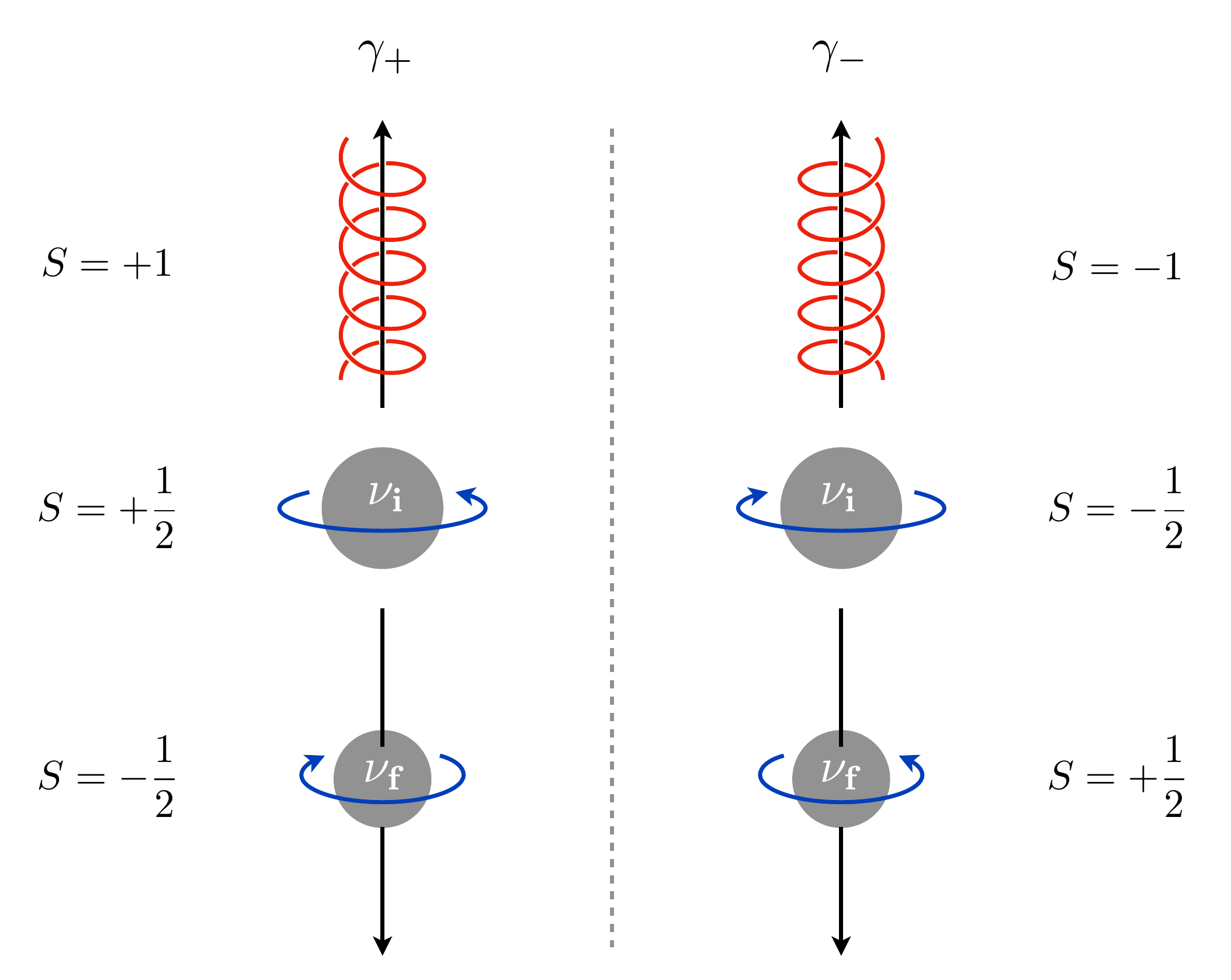}
\caption{\label{fig:radiative_decay} Polarisation for neutrino radiative decay in the rest frame. }
\end{figure}

To simplify the derivation, we prefer to work in the rest frame. Frame-independent results can be obtained straightforwardly from this case.  In the rest frame,  the initial sterile neutrino $\nu_\ini$ is at rest $p_\ini^\mu=(m_\ini,0,0,0)^T$, and the photon is released in the $+z$ direction with momentum $q^\mu=(q,0,0,q)^T$. Conservation of momentum requires $p_\fin^\mu = (E_\fin, 0,0,-q)^T$ with $q=(m_\ini^2-m_\fin^2)/(2m_\ini)$ and $E_\fin=(m_\ini^2+m_\fin^2)/(2m_\ini)$. In this frame, $S$ denotes spin along the $+z$ direction i.e. $S_z$, which takes values $\pm\frac{1}{2}$. This geometry is shown in Fig.~\ref{fig:radiative_decay}. 

The angular momentum along the $z$ direction is conserved $S_z(\nu_\ini) = S_z(\nu_\fin) + S_z(\gamma)$. For a fermion, $S_z= \pm 1/2$ and for a massless photon, $S_z= \pm1$. Given the initial state $\nu_\ini$ with spin $S_z(\nu_\ini)=+1/2 (-1/2)$, the only solution for spins in final states is $S_z(\nu_\fin)=-1/2 (+1/2)$ and $S_z(\gamma)=+1 (-1)$. In other words, the released photon is the right-handed $\gamma_+$ (left-handed $\gamma_-$). 

For the photon moving in the $+z$ direction, the polarisation vectors are as defined in \cite{Peskin:1995ev}
\begin{eqnarray} \label{eq:polarisation_vectors}
\varepsilon_+^\mu = \frac{1}{\sqrt{2}} (0,1,i,0)\,,\quad
\varepsilon_-^\mu = \frac{1}{\sqrt{2}} (0,1,-i,0) 
\end{eqnarray}
correspond to spin $S_z=+1$ and $-1$, respectively.\footnote{Here we apply the convention in the textbook \cite{Peskin:1995ev}. The definition of $\epsilon_+$ in this convention has a sign difference from the one shown in \cite{Boehm:2017nrl}. Using the convention in \cite{Boehm:2017nrl} leads to a sign difference for $i \mathcal{M}(\nu_{\ini,+\frac{1}{2}} \to \nu_{\fin,-\frac{1}{2}} + \gamma_+)$ and $i \mathcal{M}(\bar{\nu}_{\ini,+\frac{1}{2}} \to \bar{\nu}_{\fin,-\frac{1}{2}} + \gamma_+)$ in Eqs.~\eqref{eq:amplitudes_1} and \eqref{eq:amplitudes_2} and $i \mathcal{M}^{\text{M}}(\nu_{\ini,+\frac{1}{2}} \to \nu_{\fin,-\frac{1}{2}} + \gamma_+)$ in Eq.~\eqref{eq:amplitude_Majorana}.} 

In this frame, for the neutrino $\nu_\fin$ moving in the $-z$ direction, the spinors $u_S(p)$ and $v_S(p)$ with spin $\pm\frac{1}{2}$ are simplified to 
\begin{eqnarray} \label{eq:spinor}
u_{+\frac{1}{2}}(p_\fin) &=& \begin{pmatrix}
\sqrt{E+q}~ \xi_{+\frac{1}{2}}  \\
\sqrt{E-q}~ \xi_{+\frac{1}{2}}
\end{pmatrix} \,, \quad \;\;\;
u_{-\frac{1}{2}}(p_\fin) = \begin{pmatrix}
\sqrt{E-q}~ \xi_{-\frac{1}{2}}  \\
\sqrt{E+q}~ \xi_{-\frac{1}{2}}
\end{pmatrix} \,, \quad 
\nonumber\\
v_{+\frac{1}{2}}(p_\fin) &=& \begin{pmatrix}
\sqrt{E+q}~ \eta_{+\frac{1}{2}}  \\
-\sqrt{E-q}~ \eta_{+\frac{1}{2}}
\end{pmatrix} \,, \quad 
v_{-\frac{1}{2}}(p_\fin) = \begin{pmatrix}
\sqrt{E-q}~ \eta_{-\frac{1}{2}}  \\
-\sqrt{E+q}~ \eta_{-\frac{1}{2}}
\end{pmatrix} \,, \quad 
\end{eqnarray}
 with
\begin{eqnarray}
\xi_{+\frac{1}{2}}  = \eta_{-\frac{1}{2}}  = \begin{pmatrix}
1  \\ 0
\end{pmatrix} \,, \quad
\xi_{-\frac{1}{2}}  = \eta_{+\frac{1}{2}}  = \begin{pmatrix}
0  \\ 1
\end{pmatrix} \,.
\end{eqnarray}
In the massless case, $u_{+\frac{1}{2}}$ and $u_{-\frac{1}{2}}$ are purely left- and right-handed respectively (because we have assumed $\nu_\fin$ is moving in the $-z$ direction). Spinors for initial neutrino $\nu_\ini$ and antineutrino $\bar{\nu}_\ini$ are given by
\begin{eqnarray}
u_{+\frac{1}{2}}(p_\ini) &=& \sqrt{E} \begin{pmatrix}
\xi_{+\frac{1}{2}}  \\
\xi_{+\frac{1}{2}}
\end{pmatrix} \,, \quad \;\;\;
u_{-\frac{1}{2}}(p_\ini) = \sqrt{E} \begin{pmatrix}
\xi_{-\frac{1}{2}}  \\
\xi_{-\frac{1}{2}}
\end{pmatrix} \,, \quad 
\nonumber\\
v_{+\frac{1}{2}}(p_\ini) &=& \sqrt{E} \begin{pmatrix}
\eta_{+\frac{1}{2}}  \\
-\eta_{+\frac{1}{2}}
\end{pmatrix} \,, \quad 
v_{-\frac{1}{2}}(p_\ini) = \sqrt{E} \begin{pmatrix}
\eta_{-\frac{1}{2}}  \\
-\eta_{-\frac{1}{2}}
\end{pmatrix} \,, 
\end{eqnarray}
The amplitudes with definite spins in the initial and final states are then given by
\begin{eqnarray} \label{eq:amplitudes_spin}
\mathcal{M}(\nu_{\ini,+\frac{1}{2}} \to \nu_{\fin,-\frac{1}{2}} + \gamma_+) &=& +\sqrt{2} f^{\text{L}}_{\fin \ini } (m_\ini^2 - m_\fin^2) \,, \nonumber\\
\mathcal{M}(\nu_{\ini,-\frac{1}{2}} \to \nu_{\fin,+\frac{1}{2}} + \gamma_-) &=& - \sqrt{2} f^{\text{R}}_{\fin \ini } (m_\ini^2 - m_\fin^2) \,, \nonumber\\
\mathcal{M}(\bar{\nu}_{\ini,+\frac{1}{2}} \to \bar{\nu}_{\fin,-\frac{1}{2}} + \gamma_+) &=& -\sqrt{2} \bar{f}^{\text{L}}_{\ini \fin} (m_\ini^2 - m_\fin^2) \,, \nonumber\\
\mathcal{M}(\bar{\nu}_{\ini,-\frac{1}{2}} \to \bar{\nu}_{\fin,+\frac{1}{2}} + \gamma_-) &=& +\sqrt{2} \bar{f}^{\text{R}}_{\ini \fin} (m_\ini^2 - m_\fin^2) \,, 
\end{eqnarray}
Here, $\nu_{\ini,+\frac{1}{2}} \to \nu_{\fin,-\frac{1}{2}} + \gamma_+$ and $\bar{\nu}_{\ini,-\frac{1}{2}} \to \bar{\nu}_{\fin,+\frac{1}{2}} + \gamma_-$ are $CP$ conjugates, while $\nu_{\ini,-\frac{1}{2}} \to \nu_{\fin,+\frac{1}{2}} + \gamma_-$ and $\bar{\nu}_{\ini,+\frac{1}{2}} \to \bar{\nu}_{\fin,-\frac{1}{2}} + \gamma_+$ are $CP$ conjugates. 
The other channels have vanishing amplitudes, consistent with angular momentum conservation. 

We can generalise the result in Eq.~\eqref{eq:amplitudes_spin} to any inertial reference frame via spatial rotations and Lorentz boosts. These transformations change spins for fermions but leave photon polarisation invariant. Eventually, we obtain the Lorentz-invariant amplitudes $\mathcal{M}(\nu_\ini \to \nu_\fin + \gamma_\pm)$ and $\mathcal{M}(\bar{\nu}_\ini \to \bar{\nu}_\fin + \gamma_\pm)$ taking the same result as Eq.~\eqref{eq:amplitudes_spin} in any reference frame. Using the $CPT$-invariance property, namely, $\bar{f}^{\text{R,L}}_{\ini \fin} = -f^{\text{R,L}}_{\ini \fin}$, we eventually arrive at Eqs.~\eqref{eq:amplitudes_1} and \eqref{eq:amplitudes_2}. 
These are the most general results independent of either particle model or reference frame.

\section{Derivation of imaginary parts of the loop integrals\label{sec:imaginary_detail}}

The two NP contributions to the sterile neutrino radiative decay given by the new proposed interactions are shown in Fig.~\ref{fig:loop_new_ints}. In order to compute their respective matrix elements, we use the couplings of the new particles $\phi$ and $\psi$ with neutrinos and sterile neutrinos shown in Section~\ref{sec:new_int}. 

In general, we have
\begin{equation}
    i\mathcal{M}(\nu_s\to\nu_i+\gamma_\pm)=i\overline{u}(p_i)\Gamma^\mu_{is}(q^2) u(p_s)\varepsilon^*_{\pm,\mu}(q)
\end{equation}  
and the matrix elements for each loop contribution, $\mathcal{M}_j\equiv\mathcal{M}_j(\nu_s\to\nu_i+\gamma_\pm)$,  shown in Fig.~\ref{fig:loop_new_ints} take the form
\begin{eqnarray}
&i\mathcal{M}_1&=  -Qe  \lambda_s \lambda_i^* \int \frac{d^4 k}{(2\pi)^4}\frac{\overline{u}(p_i)P_\text{R} (\slashed{k}+m_\psi)(p_1-p_2)^\mu P_\text{L}  u(p_s)\varepsilon^*_{\pm,\mu}(q)}{(k^2-m_\psi^2+i\epsilon)((k-p_s)^2-m_\phi^2+i\epsilon)((k-p_i)^2-m_\phi^2+i\epsilon)} \,,\nonumber\\
&i\mathcal{M}_2&= +Qe  \lambda_s \lambda_i^* \int \frac{d^4 k}{(2\pi)^4}\frac{\overline{u}(p_i)P_\text{R} (\slashed{k}'+m_\psi)\gamma^\mu(\slashed{k}+m_\psi)P_\text{L}  u(p_s)\varepsilon^*_{\pm,\mu}(q)}{((k-p_s)^2-m_\phi^2+i\epsilon)(k'^2-m_\psi^2+i\epsilon)(k^2-m_\psi^2+i\epsilon)} \,.
\end{eqnarray}
Due to the projection operators, the matrix elements reduce to
\begin{eqnarray}\label{eq:matrix_element_LL}
&i\mathcal{M}_1&= - Qe  \lambda_s \lambda_i^* \int \frac{d^4 k}{(2\pi)^4}\frac{\overline{u}(p_i)\slashed{k}(p_1-p_2)^\mu P_\text{L}  u(p_s)\varepsilon^*_{\pm,\mu}(q)}{(k^2-m_\psi^2+i\epsilon)((k-p_s)^2-m_\phi^2+i\epsilon)((k-p_i)^2-m_\phi^2+i\epsilon)} \,\nonumber\\
&i\mathcal{M}_2&= +Qe  \lambda_s \lambda_i^* \int \frac{d^4 k}{(2\pi)^4}\frac{\overline{u}(p_i)\slashed{k}'\gamma^\mu\slashed{k}P_\text{L}  u(p_s)\varepsilon^*_{\pm,\mu}(q)}{((k-p_s)^2-m_\phi^2+i\epsilon)(k'^2-m_\psi^2+i\epsilon)(k^2-m_\psi^2+i\epsilon)} \,.
\end{eqnarray}
In order to perform  dimensional regularisation to Eq.~\eqref{eq:matrix_element_LL}, we must substitute the denominator with the relevant Feynman parameters, therefore, we perform the loop momentum shifts $\ell=k-\left(x p_s + z p_i\right)$ and $\ell=k-\left(x p_s + z q\right)$ for the two diagrams respectively. This leads to

\begin{align}\label{eq:FP_matrix_element_LL}
i\mathcal{M}_1= - Qe  \lambda_s \lambda_i^* &\int \frac{d^d \ell}{(2\pi)^d}\int dxdydz\delta(x+y+z-1)\times\nonumber\\
&\times\frac{\overline{u}(p_i)[-2\ell^\mu\slashed{\ell}+(p_s+p_i)^\mu(\slashed{p}_sy+\slashed{p_i}z)-2(p_sy+p_iz)^\mu(\slashed{p}_sy+\slashed{p_i}z)]P_\text{L}  u(p_s)\varepsilon^*_{\pm,\mu}(q)}{(\ell^2-\Delta_{\phi\psi}(x,y,z))^3} \,, \nonumber\\
i\mathcal{M}_2= + Qe  \lambda_s \lambda_i^*&\int \frac{d^d \ell}{(2\pi)^d}\int dxdydz\delta(x+y+z-1)\times\nonumber\\
&\times\frac{\overline{u}(p_i)[\slashed{\ell}\gamma^\mu\slashed{\ell}+(\slashed{q}(z-1)+\slashed{p}_sx)\gamma^\mu(\slashed{q}z+\slashed{p}_sx)]P_\text{L}  u(p_s)\varepsilon^*_{\pm,\mu}(q)}{(\ell^2-\Delta_{\psi\phi}(x,y,z))^3}\,,
\end{align}
where $\Delta_{\phi\psi}(x,y,z)$ and $\Delta_{\psi\phi}(x,y,z)$ have been defined in Eq.~\eqref{eq:Deltaxyz}. We ignore linear terms of $\ell$ since these terms vanish after integration. We use the following results from \cite{Peskin:1995ev} for $d$-dimensional integrals over $\ell$ in Minkowski space
\begin{eqnarray}
\int \frac{d^d \ell}{(2\pi)^d}\frac{1}{(\ell^2-\Delta)^n}&=&\frac{(-1)^n }{(4\pi)^{d/2}}\frac{\Gamma(n-d/2)}{\Gamma(n)}\left(\frac{1}{\Delta}\right)^{n-\frac{d}{2}}\nonumber\\
\int \frac{d^d \ell}{(2\pi)^d}\frac{\ell^\alpha\ell^\beta}{(\ell^2-\Delta)^n}&=&i\frac{(-1)^{n-1}}{(4\pi)^{d/2}}\frac{g^{\alpha\beta}}{2}\frac{\Gamma(n-d/2-1)}{\Gamma(n)}\left(\frac{1}{\Delta}\right)^{n-\frac{d}{2}-1}.
\end{eqnarray}

After dimensional regularisation, we set $d=4-\epsilon$, therefore the amplitudes acquire the following general form
\begin{align}
i\mathcal{M}_1= \frac{- iQe  \lambda_s \lambda_i^*}{(4\pi)^2}&\int dxdydz\delta(x+y+z-1)\overline{u}(p_i)\left[\left(-\frac{2}{\epsilon}+\log\frac{\Delta_{\phi\psi}(x,y,z)}{4\pi}+\gamma_\epsilon+\mathcal{O}(\epsilon)\right)\gamma^\mu\right.\nonumber\\[.2cm]
&\left.-\frac{(p_s+p_i)^\mu(\slashed{p}_sy+\slashed{p}_iz)-2(p_sy+p_iz)^\mu(\slashed{p}_sy+\slashed{p}_i z)] }{\Delta_{\phi\psi}(x,y,z)}\right]P_\text{L}  u(p_s)\varepsilon^*_{\pm,\mu}(q) \,, \nonumber\\
i\mathcal{M}_2= \frac{+ iQe  \lambda_s \lambda_i^*}{(4\pi)^2} &\int dxdydz\delta(x+y+z-1)\overline{u}(p_i)\left[\left(-\frac{2}{\epsilon}+1+\log\frac{\Delta_{\psi\phi}(x,y,z)}{4\pi}+\gamma_\epsilon+\mathcal{O}(\epsilon)\right)\gamma^\mu\right.\nonumber\\[.2cm]
&\left.-\frac{(\slashed{q}(z-1)+\slashed{p}_sx)\gamma^\mu(\slashed{q}z+\slashed{p}_sx)}{\Delta_{\psi\phi}(x,y,z)}\right]P_\text{L}  u(p_s)\varepsilon^*_{\pm,\mu}(q).
\end{align}
We simplify the above expressions by making use of the following identities
\begin{eqnarray}
\overline{u}(p_i)(p_s+p_i)^\mu P_\text{L}  u(p_s)&=&\overline{u}(p_i)[\gamma^\mu(m_s P_\text{R} +m_i P_\text{L} )+i\sigma^{\mu\nu}q_\nu P_\text{L} ]  u(p_s) \,,\nonumber\\
\overline{u}(p_i)(\slashed{p}_s+\slashed{p}_i)\gamma^\mu P_\text{L}  u(p_s)&=&\overline{u}(p_i)[2m_i\gamma^\mu P_\text{L} +i\sigma^{\mu\nu}q_\nu P_\text{L} +q^\mu P_\text{L} ]  u(p_s) \,,\nonumber\\
\overline{u}(p_i)\gamma^\mu(\slashed{p}_s+\slashed{p}_i) P_\text{L}  u(p_s)&=&\overline{u}(p_i)[2m_s\gamma^\mu P_\text{R} +i\sigma^{\mu\nu}q_\nu P_\text{L} -q^\mu P_\text{L} ]  u(p_s) \,.
\end{eqnarray}
Finally, applying the Ward identity $q^\mu \mathcal{M}_\mu=0$ and ignoring terms proportional to $\gamma^\mu$, since these are simply vertex corrections to the overall electric charge,\footnote{Notice that when both contributions are added the divergent terms cancel out.} we only need to consider the tensor-like terms within $\Gamma_{is}^\mu$ to determine the form factor resulting from these diagrams. These are given by
\begin{eqnarray}\label{eq:effective_vertex}
&\Gamma_{is,1}^\mu&= -\frac{ Qe  \lambda_s \lambda_i^*}{(4\pi)^2}i \sigma^{\mu\nu}q_\nu\int_0^1\dx\dy\dz\delta(x+y+z-1)\frac{(m_s yP_\text{R} + m_i z P_\text{L} )}{\Delta_{\phi\psi}(x,y,z)}\nonumber\\
&\Gamma_{is,2}^\mu&= +\frac{ Qe  \lambda_s \lambda_i^*}{(4\pi)^2}i \sigma^{\mu\nu}q_\nu\int_0^1\dx\dy\dz\delta(x+y+z-1)\frac{(m_s xyP_\text{R} +m_i xz P_\text{L} )}{\Delta_{\psi\phi}(x,y,z)}\,.
\end{eqnarray}

Setting $m_i\to0$ for the active neutrino mass in Eq.~\eqref{eq:effective_vertex} and integrating over $z$ yields  
\begin{eqnarray}
&\Gamma_{is,1}^\mu&= \frac{ C_1}{(4\pi)^2}i \sigma^{\mu\nu}q_\nu\int_0^1 \int_0^{1-y} \dx\dy\frac{m_s yP_\text{R} }{m_\phi^2(1-x)+x m_\psi^2-x y m_s^2}\nonumber\\
&\Gamma_{is,2}^\mu&=\frac{ C_2}{(4\pi)^2}i\sigma^{\mu\nu}q_\nu\int_0^1\int_0^{1-y}\dx\dy \frac{m_s x yP_\text{R} }{m_\psi^2(1-x)+x m_\phi^2- x y m_s^2}\,.
\end{eqnarray}
From these last expressions, we can identify the factors $K^{\text{L}}_{1,2}$ and $K^{\text{R}}_{1,2}$ given in Eq.~\eqref{eq:form_f1} and then integrate over the remaining Feynman parameters $x$ and $y$ as shown in Eq.~ \eqref{eq:Kzx}.

\end{document}